\documentclass[aps,prb,twocolumn,showpacs,amsmath,amssymb]{revtex4-1}
\usepackage{graphicx}
\usepackage{color}
\usepackage[colorlinks,plainpages=false,linkcolor=black,urlcolor=blue,citecolor=blue,pdfpagemode=UseNone,pdfstartview=FitH]{hyperref}
\usepackage[T1]{fontenc}
\usepackage[ansinew]{inputenc}
\usepackage[english]{babel}
\usepackage[left=2.5cm,right=2.5cm,top=2cm,bottom=3.0cm,a4paper]{geometry}
\usepackage{tabularx}
\usepackage{color}
\usepackage{graphicx}
\usepackage{dcolumn}
\usepackage{bm}

\newcommand{\bea}{\begin{eqnarray}}
\newcommand{\eea}{\end{eqnarray}}
\newcommand{\beq}{\begin{equation}}
\newcommand{\eeq}{\end{equation}}
\newcommand{\bd}{\begin{displaymath}}
\newcommand{\ed}{\end{displaymath}}
\newcommand{\ba}{\begin{array}}
\newcommand{\ea}{\end{array}}
\newcommand{\bi}{\begin{itemize}}
\newcommand{\ei}{\end{itemize}}
\newcommand{\bc}{\begin{center}}
\newcommand{\ec}{\end{center}}
\newcommand{\bfl}{\begin{flushleft}}
\newcommand{\efl}{\end{flushleft}}
\newcommand{\bfr}{\begin{flushright}}
\newcommand{\efr}{\end{flushright}}

\def\6{\partial}

\def\={\!\!\!&=&\!\!\!}
\def\+{\!\!\!&&\!\!\!+~}
\def\-{\!\!\!&&\!\!\!-~}

\newcolumntype{C}{@{}>{\hspace*{5mm}}c}

\newcommand{\sig}{\sigma}
\newcommand{\sigb}{\bar{\sigma}}
\newcommand{\sigs}{\sigma^\prime}

\newcommand{\yps}{\upsilon}
\newcommand{\RN}[1]{\text{\uppercase\expandafter{\romannumeral #1\relax}}}

\newcommand{\blue}{\textcolor{blue} }
\newcommand{\be}{\begin{equation}}
\newcommand{\ee}{\end{equation}}

\def\6{\partial}

\def\={\!\!\!&=&\!\!\!}
\def\+{\!\!\!&&\!\!\!+~}
\def\-{\!\!\!&&\!\!\!-~}

\begin{document}

\title[]{Superconductivity from repulsion in LiFeAs: novel $s$-wave symmetry and potential time-reversal symmetry breaking}
\author {F. Ahn$^{1}$}
\author{I. Eremin$^{1}$}
\email{Ilya.Eremin@rub.de}
\author {J. Knolle$^{2}$}
\author{V.B. Zabolotnyy$^{3}$}
\author{S.V. Borisenko$^{3}$}
\author{B. B\"uchner$^{3}$}
\author{A.V. Chubukov$^{4}$}
\affiliation {$^1$Institut f\"ur Theoretische Physik III, Ruhr-Universit\"at Bochum, D-44801 Bochum, Germany}
\affiliation{$^2$Max Planck Institute for the Physics of Complex Systems, D-01187 Dresden, Germany}
\affiliation{$^3$Leibniz-Institut f\"ur Festk\"orper- und Werkstoffforschung Dresden, P.O. Box 270116, D-01171 Dresden,Germany}
\affiliation{$^4$Department of Physics, University of Wisconsin-Madison, Madison,Wisconsin 53706, USA}

\begin{abstract}
We analyze the structure of the pairing interaction and superconducting gap in LiFeAs by decomposing the pairing interaction for various $k_z$ cuts into $s-$ and $d$-wave components and by studying the leading superconducting instabilities. We use the ten orbital tight-binding model, derived from {\it ab-initio} LDA calculations with hopping parameters extracted from the fit to ARPES experiments.
We find that the pairing interaction almost decouples  between two subsets, one consists of the outer hole pocket and two electron pockets, which are quasi-2D and are made largely out of $d_{xy}$ orbital, and the other consists of the two inner hole pockets, which are quasi-3D and are made mostly out of $d_{xz}$ and $d_{yz}$ orbitals. Furthermore, the bare inter-pocket and intra-pocket interactions within each subset are nearly equal. In this situation, small changes in the intra-pocket and inter-pocket interactions due to renormalizations by high-energy fermions give rise to a variety of different gap structures. We focus on $s-$wave pairing which, as experiments show, is the most likely pairing symmetry in LiFeAs.  We find four different configurations of the $s-$wave gap immediately below $T_c$: the one in which superconducting gap changes sign between two inner hole pockets and between the outer hole pocket and two electron pockets,  the one in which the gap changes sign between two electron pockets and three hole pockets, the one in which the gap on the outer hole pocket differs in sign from the gaps on the other four pockets, and the one in which the gaps on two inner hole pockets have one sign, and the gaps on the outer hole pockets and on electron pockets have different sign. Different $s$-wave gap configurations emerge depending on whether the renormalized interactions
increase attraction within each subset  or increase the coupling between particular components of the two subsets. We discuss the phase diagram and experimental probes to determine the structure of the superconducting gap in LiFeAs. We argue that the state with opposite sign of the gaps on the two inner hole pockets has the best overlap with ARPES data. We also argue that at low $T$, the system may enter into a "mixed" $s+is$ state, in which the phases of the gaps on different pockets differ by less than $\pi$  and time-reversal symmetry is spontaneously broken.
\end{abstract}

\date{\today}


\maketitle

\section{Introduction}\label{sec:1_introduction}
The relation between unconventional superconductivity, the electronic structure and correlations in multi-band materials is one of the most interesting topics in modern  studies of correlated electrons. The role of electronic correlations for superconductivity was extensively discussed after the discovery of superconductivity in the cuprates, for which the most natural mechanism for $d-$wave pairing is the exchange of antiferromagnetic spin fluctuations~\cite{scalapino}. Multi-band aspects have been analyzed in connection with superconductivity in MgB$_2$.\cite{mazin_2001}
The interplay between the multi-band electronic structure and magnetism has been extensively studied after the discovery of a new class of iron-based superconductors (FeSCs) -- iron pnictides and iron-chalcogenides.

Most of FeSCs contain quasi-2D hole and electron pockets, separated by $(\pi,\pi)$ in the physical 2-Fe Brillouin zone, and superconductivity emerges in a close proximity to $(\pi,\pi)$ magnetism. The widely discussed theory of superconductivity in these systems borrows concepts from the cuprates and assumes that the superconducting pairing originates from the interaction between fermions near hole and electron pockets, enhanced  by $(\pi,\pi)$ spin fluctuations\cite{kamihara,Graser09,korshunov_review,chubukov_review,Wen11}. Such  pairing mechanism yields s-wave superconductivity in which the gap changes sign between hole and electron Fermi surfaces (FSs)  -- an $s^{+-}$ state.   The fine details of $s^{+-}$ superconductivity may be quite subtle as there are at least two hole pockets and two symmetry-related electron pockets, and the gaps on these pockets differ not only by a sign but also by magnitude. Besides, the pairing interaction is commonly obtained by converting from the orbital to the band basis and generally depends on the angles along hole and electron pockets. This, together with non-circular character of Fermi surfaces,  gives rise to angular variations of the gaps,  which can be quite substantial, particularly on electron pockets. Still, the dominant property of $s^{+-}$ superconductivity is the sign change between the FS-averaged gap on hole pockets and on electron pockets. Properties of $s^{+-}$ superconductors have been studied in great detail and have been favorably compared with the large volume of data on several families of weakly/moderately doped 1111 and 122-types FeSCs~\cite{korshunov_review,chubukov_review,Wen11,paglione}.

Recently, there have been several attempts to expand the class of possible superconducting states in FeSCs beyond a "plus-minus" $s^{+-}$ state, particularly in systems which do not show strong antiferromagnetic fluctuations in the proximity of the superconducting region. One line of research focused on the interplay between s-wave and d-wave superconductivity as in most FeScs both channels are attractive~\cite{scal_njp}, another focused on systems which have only hole or only electron pockets. For the latter,  the interaction between hole and electron states is likely not the dominant pairing interaction because excitations in one of the two sets are gapped, and one has to consider an interaction between hole pockets or between electron pockets as an alternative.  In particular, in strongly hole-doped systems, like KFe$_2$As$_2$,which only have hole pockets, one proposal is that superconductivity originates from the repulsive interaction between the two $\Gamma$-centered  hole pockets, in which case  the gap must change sign between these two hole pockets~\cite{Maiti2011,maiti_korsh}. A competing proposal for the same material is  a $d_{x^2-y^2}$ state, which, according to theory~\cite{ronny_KFeAs}, is mostly concentrated on the third, largest  hole pocket. Which superconducting order develops in  KFe$_2$As$_2$ and, more generally, in Ba$_{1-x}$K$_x$Fe$_2$As$_2$, is still a subject of debate ~\cite{louis,shin, richard}, but, in any case, this example shows that superconductivity in FeSCs is not restricted to only one particular $s-$wave state, unlike in the cuprates, where the superconductivity is believed to be $d_{x^2-y^2}$-wave in all families and at all hole and electron dopings where it exists.

The possibility to have different pairing states at different doping levels of the same material lead to speculations that in the intermediate regime FeSCs may possess a superconducting state which breaks time-reversal symmetry, either $s+id$ state~\cite{hanke,khodas} or $s+is$ state.~\cite{Maiti2013,lara}
A superconducting state which breaks time-reversal symmetry has a wealth of intriguing properties and has strong potential for applications. It is then highly desirable to detect such a state in weakly/moderately doped FeSCs which contain both hole and electron pockets. Most of these systems are, however, in close proximity to antiferromagnetism, in which case the interaction between fermions near hole and electron pockets is strongly enhanced and gives rise to a conventional $s^{+-}$ state.

In this paper we argue that several novel s-wave superconducting states, including a set of $s+is$ states,
may exist in the stoichiometric LiFeAs. This material does contain both hole and electron pockets (see Fig. 1), however, in distinction to other materials, it superconducts at $T_c=17$K already at zero doping\cite{wang,chu,morozov} and shows neither static antiferromagnetic (AF) ordering nor nesting between electron and hole bands \cite{borisenko}. We argue that the following two features make this material unique with respect to superconductivity. First, due to the specific orbital content of hole and electron pockets, the system decouples into two weakly interacting subsets -- subset I contains two 3D hole pockets ($\alpha$ pockets), which exist only in a range of $k_z$ near $\pi$, and subset II contains three quasi-2D pockets -- two electron pockets ($\beta$ pockets) and one hole pocket ($\gamma$ pocket). Second, although the interactions within each subset are at least an order of magnitude stronger than the interactions between the subsets, attractive and repulsive components  within each subset are about equal and almost cancel each other.  As a result, the pairing interaction within each subset nearly vanishes.

Because of these two features, superconductivity in LiFeAs is determined by the interplay between the residual interactions within each of the two subsets and the interactions between the subsets.  Both are small if we use the bare interactions between low-energy fermions, i.e.,  the ones which are obtained by just converting the Hubbard and Hund interactions from the orbital to band basis. However, intra-subset and inter-subset interactions, which are relevant to superconductivity, generally differ from the bare ones due to renormalizations coming from high-energy fermions. For those FeSCs, in which superconductivity develops in close proximity to antiferromagnetism, the conventional recipe how to handle these renormalizations is to dress interactions by RPA-type corrections from the spin channel~\cite{scal_njp,korshunov_review}. In  LiFeAs, however, there is no evidence that magnetic fluctuations are strong~\cite{knolle12}. In
this situation one generally expects that the renormalizations of all interactions are comparable, and dressed interactions remain of the same magnitude as the bare ones. The strategy we use in this paper is to vary different couplings and check which superconducting phases appear at modest deviations from the original model.  We argue that four different $s-$wave states are possible and select the most likely state in LiFeAs by comparing the distribution of gap magnitudes on various FSs with the ARPES results.

We present the summary  of our results later in this Section, but before we list the experimental facts about LiFeAs and present a short summary of earlier theoretical works.
\begin{figure}[htbp]
 \includegraphics[width=0.45\textwidth]{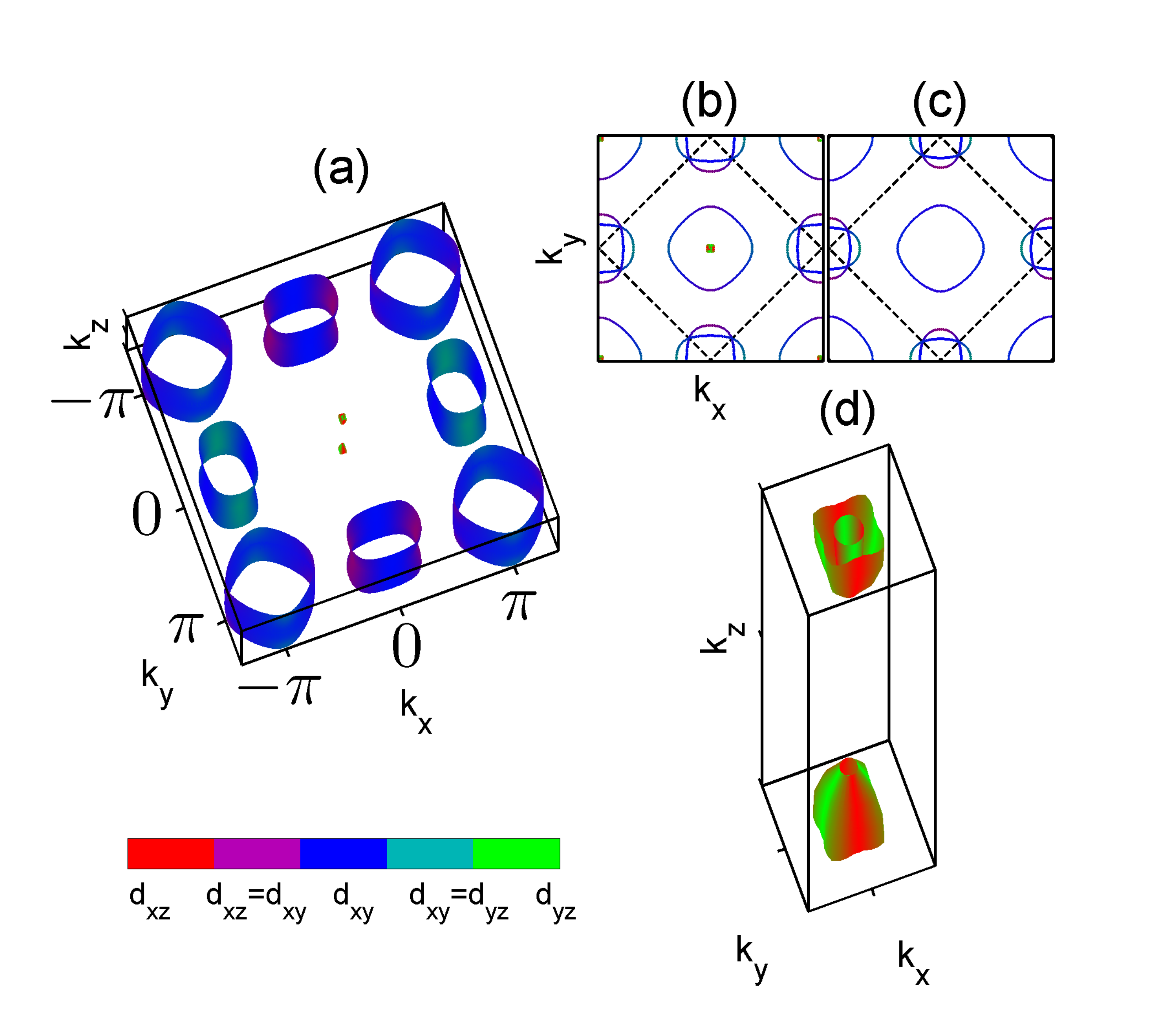}
     \caption{Fermi surface of LiFeAs as deduced from the ARPES experiments: (a) shows the three-dimensional version of the Fermi surface and (b) and (c) refer to the two-dimensional cuts at $|k_z|=\pi$ (left) and $k_z=0$, respectively. Hole pockets are located a $(0,0)$ and $(\pi,\pi)$ and electron pockets are at $(\pm\pi,0)$ and $(0,\pm\pi)$. In case of $k_z=0$, the two tiny hole pockets $h_{1,2}$ vanish just below the FS and only $h_3$ and $e_{1,2}$ remain. (d) shows the zoomed region of the first BZ around the $\Gamma-$point of the BZ with tiny $\alpha$ hole pockets.}\label{fig1}
\end{figure}

\subsection{Experimental facts about LiFeAs}

ARPES \cite{borisenko,Symmetry,umezawa} and dHvA\cite{balicas_13} data on LiFeAs show that the electronic structure does consist of three hole pockets and two electron pockets (see Fig. 1). Electron pockets ($\beta$ pockets in commonly used notations) are quasi-2D with relatively weak variation along $k_z$.  Among hole pockets, the larger one -- the $\gamma-$pocket, is also quasi-two-dimensional and is centered at $(\pi,\pi)$ in the Fe-only Brillouin zone (1Fe BZ). The other two hole pockets (the $\alpha$-pockets) are centered at $\Gamma$ in both 1Fe BZ and the physical 2Fe BZ and reach the Fermi level only at $k_z$ near $\pi$. Even there, the $\alpha$ pockets are small and it is possible that one of them remains very close but still below the Fermi energy.  At small $k_z$ the maxima of the dispersion of the two $\Gamma-$ bands are far below the Fermi level. ARPES measurements below $T_c$ show that the superconducting gap is the largest  on one of $\Gamma$-centered $\alpha-$pockets ($\Delta_\alpha \sim 6 meV$ at $k_z$  where it exists). The gaps on $\gamma$ and $\beta$ pockets are somewhat smaller, between $3$ meV and $4$ meV. The gaps do not have nodes, but do show sizeable variations along the FS: the gap on the $\gamma$ pocket has a  $\cos 4 \theta$ variation along the FS\cite{Symmetry,umezawa} and the gaps on the two electron $\beta$ pockets vary nearly as $\pm |\cos 2 \theta|$ \cite{Symmetry,umezawa}, which is expected when the hybridization between the two $\beta$ pockets is weak. Probably, the gap on the $\alpha$ pocket also has  an angular dependence, but the pocket is too small to detect the angular dependence along it  in ARPES measurements\cite{Symmetry,umezawa}.

The absence of gap nodes rules out pure non-s-wave states, like $d-$wave.  More complex states like $s+id$ or $p+ip$ are possible. However, the observed
$\cos 4 \theta$ gap variation on the $\gamma$ pocket strongly suggests that the gap is $s-$wave, like in other FeSCs with sizable hole and electron pockets. The issue, which ARPES experiments cannot resolve, is whether the $s$-wave state  is a conventional $s^{+-}$ with a positive gap on all hole pockets and a negative one on electron pockets (or vise versa), or some other $s-$wave with, {\it e.g.}, plus-minus gap between the two $\Gamma$-centered $\alpha-$pockets.
The quasiparticle interference experiments can, in principle, determine relative signs of the gaps on various FSs but the data for LiFeAs are not conclusive at the moment.\cite{allan,damascelli,hess} Like we said, a conventional $s^{+-}$ state develops if there is a strong interaction between hole and electron pockets.
The enhancement of electron-hole interaction is believed to come from magnetic fluctuations. Given that the gap on the $\alpha$ pocket is larger than that on the $\gamma$ pocket, spin fluctuations must be strong at momenta near $(\pi,\pi)$. One way to verify this experimentally is to check whether the magnetic response at or near ${\bf Q}=(\pi,\pi)$ shows a resonance peak  below $T_c$~\cite{maiti_res}. Several neutron scattering experiments on LiFeAs have been performed recently \cite{taylor,wang_dai,qureshi}, including  one study\cite{qureshi} on superconducting single crystals. An enhanced intensity at an incommensurate momentum close to $(\pi,\pi)$  has been observed, however its variation across $T_c$ was found to be too weak to draw a definite conclusion whether or not there is a resonance peak below $T_c$~\cite{knolle12}.

\subsection{Earlier theoretical works}

On the theory side, several groups analyzed superconducting order and the structure of magnetic excitations in LiFeAs using different techniques. A functional renormalization group (fRG) study found a conventional $s^{+-}-$wave superconductivity, which develops simultaneously with $(\pi,\pi)$ antiferromagnetic fluctuations\cite{platt_12}. Wang et al~\cite{wang_13} analyzed superconductivity in LiFeAs within the RPA scheme, using the ten-orbital 3D band dispersion chosen to match the FSs observed in ARPES. This group also found a conventional $s^{+-}$ superconductivity, driven by $(\pi,\pi)$  interaction between  fermions near hole and electron pockets. They calculated the ratios of the superconducting gaps at $T = T_c-0$ and found a good match with the gap ratio on the $\gamma$ and $\beta$ pockets. However, the gap on the $\alpha$ pocket turns out to be smaller than the other gaps, in disagreement with the ARPES results, which show that the $\alpha$ gap is the largest. Ummarino et al  argued\cite{ummarino_13} that one can match the ratios of all gaps in an $s^{+-}$ state if one combines magnetic fluctuations at large $q \approx {\bf Q}$ and small $q$ fluctuations, for which, they argued, the best candidate is electron-phonon interaction. Another group also found~\cite{brydon} that interactions at small momentum transfer are strong but attributed it to strong small $q$ magnetic fluctuations. A chiral $p+ip$ state driven by small $q$ magnetic fluctuations has been proposed but not analyzed in detail within a microscopic model\cite{brydon}.

Yin et al~\cite{kotliar} analyzed superconductivity in LiFeAs within the DMFT and obtained a different s-wave state, in which the (FS averaged) gaps on $\alpha$ and $\beta$ pockets have one sign, and the gap on the $\gamma$ pocket has another sign.

We did find, in some range of parameters, both the conventional $s^{+-}$ state and the s-wave state obtained by Yin et al\cite{kotliar}. For other parameters, however, we found two other s-wave states, not discussed in earlier literature on LiFeAs. In addition, we argue that $s+is$-superconductivity of one type or the other can emerge al a low $T < T_c$.

\subsection{Summary of our results}

We analyzed  superconductivity in LiFeAs using the same  ten-orbital 3D band dispersion and FSs extracted from ARPES as in Ref.\cite{wang_13}. For the interaction we used the same model as in earlier studies~\cite{scal_njp}, with Hubbard $U$ and Hund $J$ local interactions in the orbital basis. In distinction to Ref. \cite{wang_13}, however, we did not rely on the RPA procedure for a fixed choice of parameters $U$ and $J$, which are not known precisely anyway. Instead, our strategy was to consider $J/U$ as a parameter,  convert from orbital to band basis, decouple the interactions into orthogonal $s-$wave, $d_{x^2-y^2}$ and $d_{xy}$ channels, and approximate each interaction by the leading and subleading angular harmonics, consistent with the given symmetry.  Such  an approach has been previously applied to 1111 and 122-type systems~\cite{Maiti2011}, where it was found that each intra-pocket and inter-pocket interaction is well approximated by the first two  angular harmonics.  Carrying out this procedure for three different values of $J/U$, we obtained in each case the pairing Hamiltonian with a finite number of the interaction terms. We solved the BCS-type gap equations at $T=T_c$ and for each $J/U$ analyzed possible superconducting states, first for bare interactions and then by allowing different interaction components to vary modestly. We assumed, like in Ref. \cite{Maiti2011}, that varying the interaction components  mimics the effect of static renormalizations by high-energy fermions. In particular, increasing the interactions between fermions on hole and electron pockets mimics the effect of antiferromagnetic spin fluctuations.
\begin{figure}[htbp]
\includegraphics[width=0.4\textwidth]{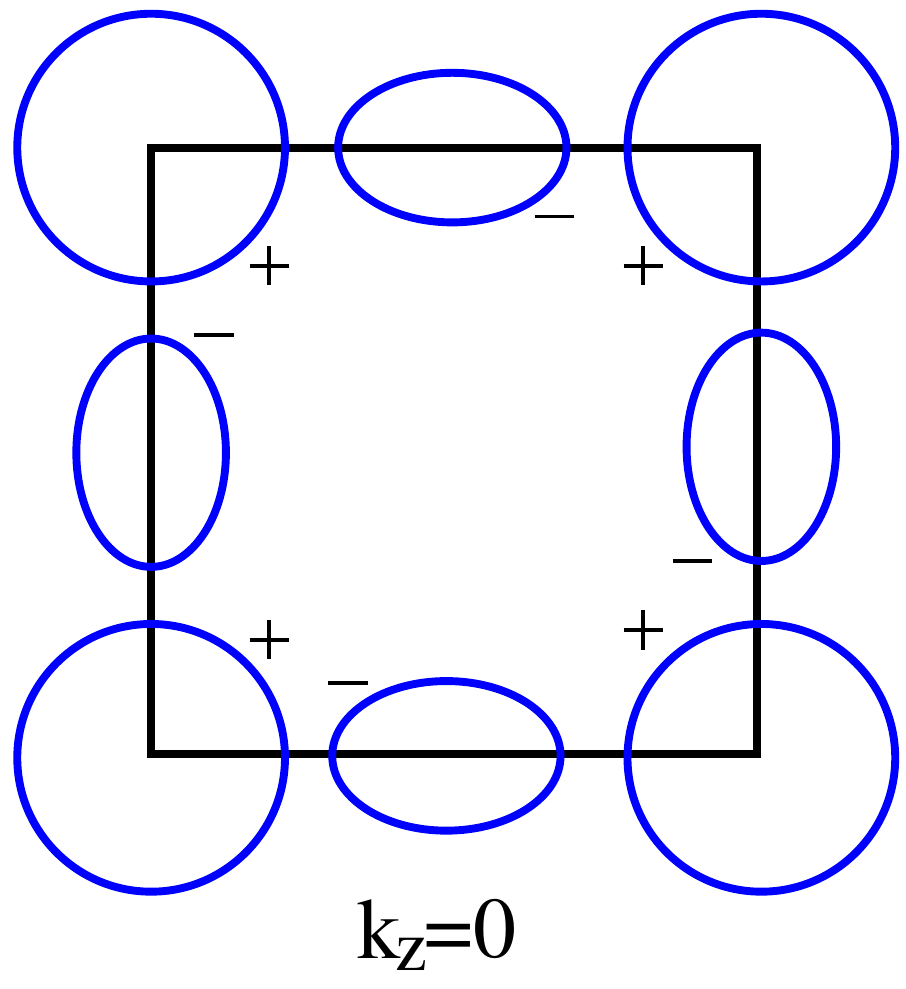}
\caption{The gaps on the $\gamma$ hole pocket and two $\beta$ electron pockets in the $s^{+-}-$ state for $k_z=0$. The pockets are presented in the 1Fe zone, with the $\gamma$ pocket centered at $(\pi,\pi)$.}\label{fig:0}
\end{figure}

We solved for the pairing in two 2D cross-sections. One is at $k_z =0$, when only subset II is present (the $\gamma$ pocket and the two $\beta$ pockets), another at $k_z = \pi$, when both subset I and subset II are present. We assume that the pairing potential does not extend over a wide range of $k_z$, in which case each cross-section can be analyzed independently. We note, however, that this does not imply that superconducting gaps are independent on $k_z$, and for one state (state D, see below) this dependence is strong.

We begin by discussing our results for $s-$wave pairing.
We found that, near $k_z=0$, where only $\gamma$ and $\beta$ pockets are present, the only possible s-wave superconducting state is a conventional $s^{+-}$, in which FS-averaged gap on the $\gamma$ pocket has opposite sign to to FS-averaged gaps on the two $\beta-$pockets (see Fig. \ref{fig:0}).  For bare  interactions, the eigenvalue in $s^{+-}$ channel is close to zero for all ratios of $J/U$ which we studied. Once inter-pocket interaction get enhanced, $s^{+-}$
superconductivity becomes robust.

The angle dependencies  of the gaps are $a + b \cos 4 \theta$ on the $\gamma$-pocket and, approximately, $c \pm d |\cos 2 \theta|$ on the two $\beta$ pockets.
The angle dependencies come from two sources: one is the angular dependence of the interactions, another is the non-circular form of the FSs.  We analyzed
only the angle dependence coming from the interactions (i.e., we approximated FSs by circles).  On the $\beta$ pockets, we  found that the relative magnitude of the $|\cos 2 \theta|$ is comparable to that seen in the experiments\cite{Symmetry}. On the $\gamma$ pocket, we found that $\cos 4 \theta$ dependence of the gap, caused by the interaction is very weak and, moreover, has the wrong sign. It is then likely that the observed $\cos 4 \theta$ variation of the gap on the $\gamma$ pocket is caused not by the angle dependence of the interaction but by the $\cos 4 \theta$ FS anisotropy, as Ref.[\onlinecite{maiti_korsh}] suggested.

For $k_z \leq \pi$, all five pockets are present and one has to include the interactions between all pockets. We found that the matrix gap equation nearly
decouples between subsets I and II. We argue that this is a natural consequence of the specific orbital content of the FSs in LiFeAs. Namely, as Fig.\ref{fig1} shows, the $\gamma$ pocket and the two $\beta$ pockets are predominantly made out of $d_{xy}$ orbitals, with very little admixture of $d_{xz}$  and $d_{yz}$ orbitals. On the contrary, the two $\alpha$ pockets are made out of only  $d_{xz}$ and $d_{yz}$ orbitals.   As a result, the interaction between the subset of two $\alpha$ pockets and the subset of  $\gamma$ and $\beta$ pockets  is much weaker than inter-pocket and intra-pocket interactions within each of the two subsets. Number-wise, for all three values of $J/U$ which we used,  inter-subset interactions are one order of magnitude weaker than intra-subset interactions.
\begin{widetext}

\begin{figure}[t!]
\includegraphics[width=0.7\textwidth]{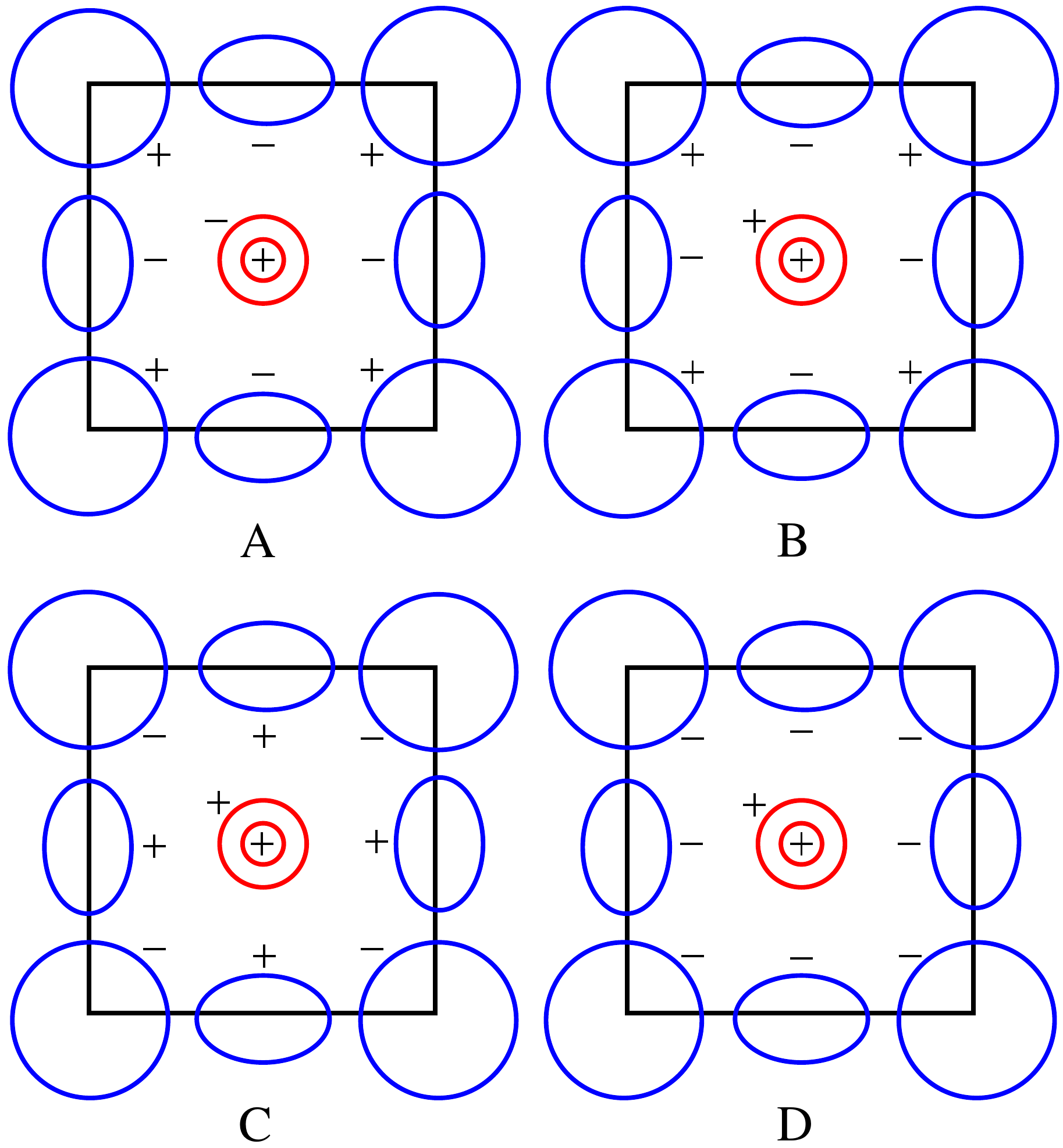}
     \caption{Four different  $s-$wave solutions at  $k_z=\leq \pi$.  Red and blue colors correspond to FSs belonging to subsets I and II, respectively.
      The ''conventional" $s_{+-}$ state is the state B. The three other states, A, C, and D, are specific to the case of two weakly coupled subsets.
      The state C has been considered in Ref. \cite{kotliar}.  Each of the four states has the largest eigenvalue in a finite range of model parameters.
      The gap structure of the state A has the best overlap with the ARPES data.}\label{fig:1}
 \end{figure}

 \end{widetext}
All intra-subset interactions are positive, in which case the only possible pairing state within each subset is $s^{+-}$. For subset I this implies a sign change between the gaps on the two $\alpha$ pockets, for subset II this implies the same $s^{+-}$ state as at $k_z=0$. For bare interactions,  we found, for all three values of $J/U$,  that the eigenvalues for $s^{+-}$ solutions are very small  in both subsets because repulsive (intra-pocket) and attractive (inter-pocket) interactions  are almost identical. Once we increase inter-pocket interactions within each subset, $s^{+-}$ pairing  becomes solid. The resulting superconducting state is the state A shown in Fig. \ref{fig:1}a. In this state, the gap structure in the subset II is the same as at $k_z=0$, hence one should not expect qualitative changes  with $k_z$, although the magnitude of the gap on $\gamma$ and $\beta$ pockets does change between $k_z =0$ and $k_z = \pi$. The gap in the subset I changes sign between the two $\alpha$ pockets, i.e., on one pocket the gap is of the same sign as on the $\gamma$  pocket, and on the other it is opposite. It turns out that the gap on a larger of the two $\alpha$ pockets (the one which is seen in ARPES) is opposite in sign to the gap on the $\gamma$ pocket.

Once we increase the strength of the coupling between subsets (either instead or in addition to increasing inter-pocket interactions within subsets), we find
that the state A evolves and eventually the gap on one of the hole pockets changes sign and the system ends up  in one of three other $s-$wave states.
One of these new states, which we label B  in Fig. \ref{fig:1}b, is a conventional $s^{+-}$ state in which the gap has one sign on all three hole pockets and another sign on the two electron pockets. This state emerges at small enough $J/U$, when we increase the interaction between $\alpha$ and $\beta$ pockets.
For small $J/U$ the $\alpha-\beta$ interaction is repulsive, and its enhancement is expected if antiferromagnetic fluctuations are strong. Not surprisingly,
our result for stronger $\alpha-\beta$ interaction coincides with spin-fluctuation studies of Wang et al~\cite{wang_13} and
Ummarino et al~\cite{ummarino_13}.

For  larger $J/U$, the interaction between $\alpha$ and $\beta$ pockets turns out to be attractive. When this interaction gets enhanced, the system ends up in a state in which the gaps on $\alpha$ and $\beta$ pockets have the same sign.  The gap on the $\gamma$ pocket has opposite sign because of strong repulsion between $\beta$ and $\gamma$ pockets. We label this state C and show it in Fig.\ref{fig:1}c. This is the same state which Yin et al have found~\cite{kotliar}.

When $\alpha-\beta$ and $\alpha-\gamma$ interactions both get larger, and $J/U$ is small enough such that both are repulsive, the system chooses yet another $s-$wave state, in which the gaps on $\beta$ and $\gamma$ pockets have one sign, and the gaps on $\alpha$ pockets have another sign.
We label such state as D and show it in Fig.\ref{fig:1}d. To obtain the state $D$, the combined effect from $\alpha-\beta$ and $\alpha-\gamma$ interactions must overcome the repulsive interaction between $\beta$ and $\gamma$ points.

We present the phase diagrams in Figs. \ref{fig_new2} and \ref{fig_new2_1}. We emphasize that the reason why  relatively weak inter-subset interactions can significantly change the gap structure and transform the state A into one of the other three states is that, for bare interactions,  the largest eigenvalue in each of the two subsets is near zero, hence the system is quite susceptible  to small perturbations.

\begin{widetext}

 \begin{figure}[t!]
 \includegraphics[width=1.0\textwidth]{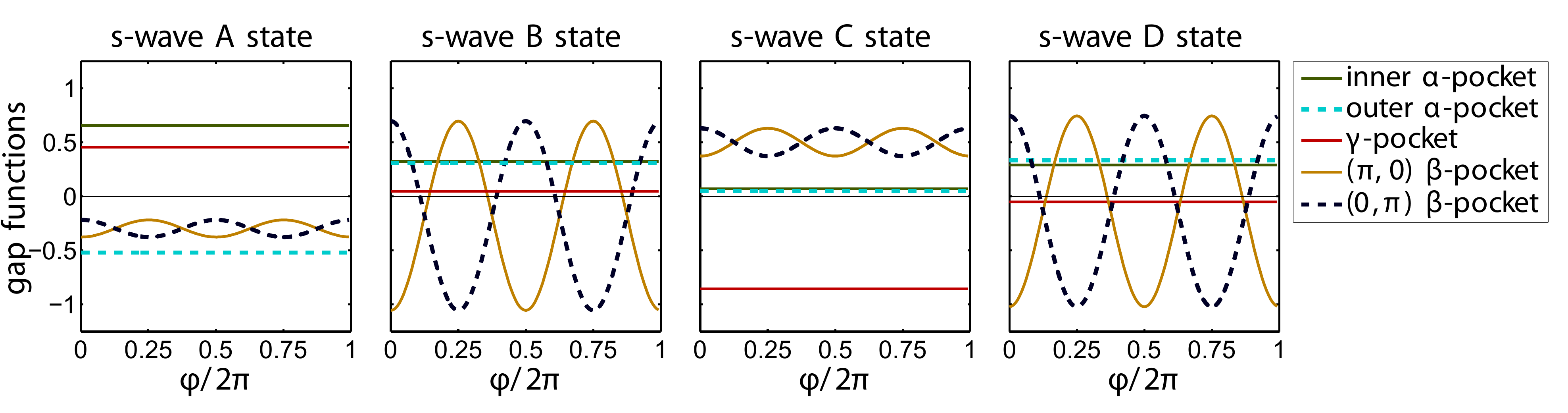}
     \caption{Distribution of the superconducting gaps on the LiFeAs Fermi surfaces for four different  $s-$wave solutions at  $k_z=\leq \pi$. The
      interaction parameters
      are shown in Fig. \ref{fig_new2_1} (b),(e). The gap structure of the state A has the best overlap with the ARPES data.}
     \label{fig:gaps}
 \end{figure}

\end{widetext}

The four s-wave states have been obtained by solving the linearized gap equation and appear right at $T_c$. The results for the gap variation along different FSs in these states are presented for some characteristic interaction parameters in Fig.\ref{fig:gaps}. We emphasize that these $s$-wave states are {\it not} orthogonal states, and  when interaction parameters change, the system gradually evolves from one state to the other. In this respect, the states A, B, C, and D are not different states in a thermodynamic sense, but rather different realizations of the s-wave gap in a multi-band superconductor. We show examples of the system evolution from state A to states B and C in Figs. \ref{fig3} and \ref{fig4}, respectively, using the paths along the lines AB and AC in Fig. \ref{fig_new2_1}. We found that the evolution is continuous at $T_c$, except for one special case.

We compared the gap structure in the states, A, B, C, and D, and found that the one which has the best agreement with ARPES data for LiFeAs is the state A. In this state, the gap on both $\alpha$ pockets turns out to be the largest out of all gaps, and the gap on the $\gamma$ pocket is slightly larger than the FS-averaged gap on the $\beta$ pockets. Both results are consistent with ARPES\cite{Symmetry}. The conventional $s^{+-}$ state (the state B) agrees less with ARPES and, at least in some range of parameters, has accidental nodes on electron pockets. The latter is inconsistent with both ARPES\cite{Symmetry} and thermodynamic data\cite{prozorovLiFeAs}, which indicate that the gaps in LiFeAs have no nodes. Still, for other parameters the B state is nodeless and we cannot rule it out as a potential superconducting state in LiFeAs. The states C  and D are less likely candidates as the state C has the largest gap on the $\gamma$ pocket, in disagreement with ARPES, and the state D has nodes on electron pockets in most of the parameter range  where this state develops at $T_c$.
We propose the experiment to verify whether the superconducting state in LiFeAs is the state A.  Namely, we solved  for $T_c (k_z)$ separately at $k_z$  near $\pi$, where both subsets are present, and at $k_z$ near zero, where only subset II is present, and found that $T_c$ is larger for $k_z$ near $\pi$. In real 3D systems there is only one $T_c$ for the whole system, which, within BCS theory, coincides with the largest  $T_c (k_z)$ in $k_z$ scans.  Below this 3D $T_c$, the gaps become non-zero for all $k_z$. However, because $T_c (k_z=\pi) > T_c (k_z=0)$, the gaps on $\gamma$ and $\beta$ pockets at $k_z \approx 0$ remain small  until the temperature gets lower than $T_c (0)$.  At the same time, the gaps at $k_z \approx    \pi$ have conventional BCS forms with $T_c = T_c (\pi)$.
This can be explicitly verified in ARPES by analyzing temperature  dependencies of the gaps in LiFeAs at various $k_z$.

We considered the range $T< T_c$ in more detail. The key result here is similar to the one found in earlier studies of superconductivity in Ba$_{1-x}$K$_x$Fe$_2$As$_2$ (Refs. \onlinecite{Maiti2013,lara}). Namely, we found  that the system evolution from the state A to other states (B,C, or D) is not necessary continuous at low enough $T$. Specifically, instead of the evolution in which the gap on one of $\alpha$ pockets reduces its magnitude to zero and then re-appears with a different sign, as it happens along the $T_c$ line (see Figs. \ref{fig3} and \ref{fig4}), the system prefers, at low enough $T$, to invert the sign of one of the $\alpha$ gaps by gradually changing the phase of the gap rather than its magnitude. This gives rise to a set of  intermediate $s+is$ states  in which the phases of the gaps on the two $\alpha$ pockets differ by less than $\pi$, i.e., time-reversal symmetry is broken. We show two examples schematically in Fig. \ref{fig:E}. We present the phase diagrams of the system behavior below $T_c$ in Fig. \ref{fig2}.

Whether or not a $s+is$ phase exists depends on parameters, and there is no guarantee that LiFeAs enters  into a state with broken time-reversal symmetry  at some $T < T_c$. Still, this is in intriguing possibility, particularly given that system parameters can, in principle, be varied by applying pressure.  The "zero-order" evidence for an $s+is$ state will  be the observation of a phase transition within the superconducting state at some $T< T_c$.

For completeness, we also analyzed possible $d-$wave superconductivity with both $d_{x^2-y^2}$ and $d_{xy}$ gap structure. We found that $d_{x^2-y^2}$ superconductivity is a strong competitor to $s-$wave for bare values of system parameters but becomes less favorable if we modify $s-$wave and $d-$wave interactions by the same amount. This is in line with the ARPES data which clearly rule out d-wave pairing in LiFeAs. The $d_{xy}$  superconductivity is even weaker.

The rest of the paper is organized as follows.
In the next section we discuss the tight-binding Hamiltonian for LiFeAs.  In Sec.~\ref{sec:3_interaction_terms} we describe our treatment of the interaction terms, decompose the pairing interactions at $k_z =0$ and $k_z = \pi$ into $s-$ and $d$-wave components, and show the near-separation into two weakly interacting subsets of two $\alpha$ pockets (subset I) and the other three pockets (subset II). In Sec.~\ref{sec:4_sc_gap} we analyze the leading superconducting instabilities in the $s-$wave channel.  We present the solutions for the bare interactions and for modified interaction parameters.  We show that the gap has one of four different structures (A,B,C,or D), and  discuss how one structure evolves into the other as we vary the interactions. In this Section we also present analytical reasoning and  compare the four gap configurations  with ARPES data. In Sec. \ref{below_tc}  we discuss the system behavior below $T_c$ and the appearance of an $s+is$ state. In Sec. \ref{d-wave} we present, for completeness, the solutions for the gap in $d_{x^2-y^2}$ and $d_{xy}$ channels. We present
our conclusions in Sec.~\ref{sec:7_conclusion}.

\section{Tight-binding Hamiltonian}
\label{sec:2_H_tb}

Tight-binding models, including up to ten bands, have been developed to describe the band structure of several families of FeSCs, including LiFeAs. \cite{Eschrig, Graser}. The need for a ten-band model for the full description of fermion hopping is due to the fact that there are five Fe d-orbitals at a given  site ($d_{xy}, d_{x^2-y^2}, d_{xz}, d_{yz}, d_{3r^2-z^2}$) and two non-equivalent positions of a pnictogen (As in our case), which is located either above or below the Fe-plane in checkerboard order.  The unit cell in the Fe plane then contains two Fe atoms, and this doubles the number of Fe orbitals required to fully describe the hopping. How important it is to use the ten-orbital model and not the five-orbital one depends, at least to a certain degree, on the structure of the superconducting gap.  The argument is that the new physical effect, which is present in the 10-orbital model, but not in the 5-orbital one, is the hybridization between the pockets separated by ${\bf Q} = (\pi,\pi)$ in the 1Fe BZ (more accurately, ${\bf Q}=(\pi,\pi,0)$ in 1111 and 111 systems, like LiFeAs, and ${\bf Q} = (\pi,\pi,\pi)$ in 122 systems). The two sets of pockets in LiFeAs, which are separated by ${\bf Q}$, are the two $\beta$ pockets (one centered at $(0,\pi)$ and another at $(\pi,0)$ in the 1Fe BZ) and $\alpha$ and $\gamma$ pockets ($\alpha$ pockets are centered at $(0,0)$ and $\gamma$ pocket is centered at $(\pi,\pi)$ in the 1Fe BZ). We argue below that $\alpha$ and $\gamma$ pockets are weakly coupled because of different orbital content, hence the hybridization between the two does not seem to be relevant.  For the two $\beta$ pockets, hybridization does matter and may even give rise to a new superconducting order~\cite{khodas},  however for this the gaps on the two $\beta$ pockets have to be of opposite sign  before hybridization. This is not the case for $s-$wave gap structure. For $s-$wave pairing, the hybridization splits the gaps on inner and outer $\beta$ pockets but it is not expected to change the physics in a qualitative way. By this reason, we neglect the hybridization and approximate the ten-orbital model by the block-diagonal form. \cite{Eschrig}
$H_0=\bigl[\begin{smallmatrix}
H^{ss}&\\
&H^{aa}
\end{smallmatrix}\bigr]$,
where each block is a $5\negthinspace\times\negthinspace5$ matrix. The block matrices $H^{ss/aa}_k$ are equivalent up to the momentum shift ${\bf Q}$. To avoid a confusion, we focus only on the $5\negthinspace\times\negthinspace5$ block in which $\alpha$ pockets are around $k_x=k_y =0$. We note, for comparison, that Wang et al~\cite{wang_13} also approximated the ten-orbital model by two non-interacting five-orbital models, but presented the FSs in both geometries in their plots.

The $5\negthinspace\times\negthinspace5$ orbital tight-binding Hamiltonian is diagonalized by a canonical transformation to
\beq
H^{aa} = \sum_{i=1}^5 \epsilon_i (k) a^\dagger_{i,k} a_{i,k}
\eeq
and the eigenvalues $\epsilon_i (k)$ describe five quasiparticle dispersions  in the band basis. The locations of hole and electron pockets are specified by $\epsilon_i (k) = \mu$.   In our calculations, we set the parameters of the tight-binding model (the elements of the $5\negthinspace\times\negthinspace5$ matrix) to match  the
low energy band structure extracted from the ARPES data\cite{wang_13}. The resulting FS's  are depicted in Fig. \ref{fig1} in the 2Fe BZ, which is obtained by folding the 1FeBZ with momentum ${\bf Q} = (\pi,\pi,0)$. The folding implies that one has to re-draw the FSs in the new BZ with momenta  ${\tilde k}_x = k_x + k_y$ and ${\tilde k}_y = k_x -k_y$, ${\tilde k}_z = k_z$. We emphasize that the folding mixes the states only within a cross-section at a given $k_z$.  This is a peculiarity of 111 electronic structure of LiFeAs in which $As$ at a given $(k_x,k_y)$  remains either above or below the nearest Fe plane for all values of $z$.  In 122 systems, the position of $As$  with respect to the Fe plane oscillates with $z$ and the folding occurs with the wavevector $(\pi,\pi,\pi)$. The colors in Fig. \ref{fig1} show the orbital content of each FS, i.e., the d-orbital component which has the largest overlap with the corresponding band operator (the largest coherence factor in the linear transformation from the orbital to band basis).

There are several prominent features in Fig. \ref{fig1}.
\begin{itemize}
\item
Like in other FeSCs, there are hole and electron pockets. Hole pockets (two $\alpha$ pockets and $\gamma$ pocket) are centered at $k_x=k_y =0$, electron pockets (the two $\beta$ pockets) are centered at $(\pi,\pi)$
\item   FSs for $\gamma$ and $\beta$ bands are quasi-2D, with rather small dispersion along $k_z$.  On the contrary, the two $\alpha$ pockets have strong $k_z$-dispersion and are only present at $k_z > 0.6 \pi$.
 As a result, at small $k_z$, the  FS in the $k_x,k_y$ plane consists of one hole and two electron pockets, while in the cross-section at $k_z$ near $\pi$, the FS consists of three hole pockets and two electron pockets.
 \item
The orbital content of the two $\alpha$ FSs is very different from that of the other three FSs.  The $\alpha$ FSs are made chiefly of $d_{xz}$ and $d_{yz}$ orbitals.  The other three FSs are made primarily of $d_{xy}$ orbital, with rather small admixture of $d_{xz}$ and $d_{yz}$ orbitals.
 \end{itemize}

These features indicate that the low-energy electronic structure of LiFeAs consists of two very different subsets. One is made out of quasi-2D $\gamma$ and $\beta$ pockets with primarily $d_{xy}$ orbital content, and the other is made out of $\alpha$ pockets, which are highly anisotropic along $k_z$ and are  made primarily out of $d_{xz}$ and $d_{yz}$ orbitals.  We will see below that this separation into the subsets holds also for the pairing interactions and gives rise to an unconventional structure of the superconducting gap in LiFeAs.

 \section{Interaction terms}
 \label{sec:3_interaction_terms}

To describe superconductivity, one needs to know the structure of the interactions. In a BCS-type treatment, which we adopt here, relevant interactions are  between fermions located right on the FSs. To obtain these interactions, one needs to convert the interaction Hamiltonian from the orbital to band basis and project it onto the FSs.
This procedure is rather straightforward and has been applied before (see e.g., Ref.\cite{Maiti2011}).  One takes as the point of departure  the model with on-site Hubbard and Hund interactions in the orbital basis, re-expresses the interaction in terms of band operators, which are linear combinations of orbital operators with coefficients obtained from the diagonalization of the tight-binding Hamiltonian, and then projects the interactions in the band basis onto the FSs. The coefficients of the transformation from the orbital to band operators (the coherence factors) depend on the position of the 3D momenta on the FS,
so in general one has to solve the full 3D gap equation.  We will adopt a more restrictive approach and solve for the pairing in two 2D cross-sections -- one at $k_z=0$, when only one subset of FSs is present, and the other at $k_z = \pi$, when  all five FSs are present.  The first cross-section is representative for all $k_z < 0.6 \pi$, the second one is representative for $0.6 \pi < k_z <\pi$.  There is a crossover  from one behavior to the other ar $k_z \sim 0.6 \pi$, and to describe it the fill 3D solution is needed. We argue, however, that the  generic structure of the pairing state at $T_c$ and the evolution of the gap structure below $T_c$ are determined chiefly by the behavior at large and small $k_z$ and depend little on the details of the crossover behavior near $k_z = 0.6 \pi$.

To solve for the pairing, we adopt the same approach as in Ref. [\onlinecite{Maiti2011}]. Namely, we separate each interaction between fermions on $i$ and $j$ FSs ($i,j =1-5$)  into $s-$wave, $d_{x^2-y^2}$, and $d_{xy}$ channels, and restrict with the leading angle momentum harmonics in each representation to make the gap equation tractable analytically and be able to follow the gap evolution upon changing the parameters. The restriction with the leading angular momentum components was termed~\cite{Maiti2011} leading angular harmonic  approximation (LAHA).  We use LAHA, solve for superconductivity and vary the parameters of the underlying model to see whether the solutions that we find are stable with respect to perturbations.

The point of  departure for our analysis is the multi-orbital local Hamiltonian, which includes on-site density-density (Hubbard) and exchange (Hund) intra-orbital and inter-orbital interactions. The Hamiltonian is given by
\begin{eqnarray}
\lefteqn{H_{\text{int}} =  \frac{U}{2}\sum_{\substack{i,\nu\\\sig}}n_{i\nu\sig}n_{i\nu\sigb}+ \frac{U^\prime}{2}\sum_{\substack{i,\nu\neq\yps\\\sig,\sigs}}n_{i\nu\sig}n_{i\yps\sigs}}&& \nonumber \\
&& - J\sum_{\substack{i,\nu\neq\yps}} S_{i\nu}\cdot S_{i\yps}+\frac{J^\prime}{2}\sum_{\substack{i,\nu\neq\yps\\\sig}}d_{i\nu\sig}^\dag d_{i\nu\sigb}^\dag d_{i\yps\sigb}d_{i\yps\sig}\;.
\end{eqnarray}
Different symbols are for the intra-orbital Hubbard interaction $U$, inter-orbital Hubbard interaction $U'$, inter-orbital
exchange $J$, and pair hopping term $J'$. We follow Refs.\cite{oles,takimoto} and assume that, to a reasonable accuracy, the interactions can be thought as originating from a single two-body term with spin rotational invariance, in which case
$J' = J$ and $U' = U - 5 J/2$ (Ref.\cite{korshunov_review}).   This leaves $U$ and $J$ as the only two parameters in the problem.  Moreover, one of these parameters, say, $U$, sets the overall magnitude of the pairing interaction, while its structure, which determines the structure of the superconducting gap, depends on the single parameter $J/U$, which we will vary.

The BCS Hamiltonian in the band basis has the form
\beq
\sum_{i,j =1}^5 \sum_{k_i, p_j} U_{i,j} ({\bf k}_{F,i}, {\bf k}_{F,_j}) \left[a^{\dagger}_{{\bf k},i} a^{\dagger}_{-{\bf k},i} a_{{\bf p},j} a_{-{\bf p},j} + h.c \right]
\label{w_1}
\eeq
where $a_{{\bf k},i}$ and $a_{{\bf p},j}$ are creation and annihilation operators in the band basis, and $U_{i,j} ({\bf k}_{F,i}, {\bf k}_{F,_j})$ are the pairing interactions projected onto the FSs labeled $i$ and $j$.  We neglected spin indices in  (\ref{w_1}) simply because the pairing states which we consider - an $s-$wave and a $d-$wave, are spin-singlet states and have the same spin structure. We do not consider a $p$-wave pairing channel.

Each $U_{i,j}$ is a linear combination of $U$ and $J$ terms dressed by the coherence factors from the transformation from orbital to band basis.
The coherence factors depend on the position of ${\bf k}$ and are not rotationally-invariant. This leads to the dependencies of the interactions on momenta along the FSs. We parametrize ${\bf k}_{F,i}$ in terms of the FS angle $\theta_i$ and count  $\theta_i$ for all $i$ as deviations from the same axis.

The LAHA approximation has been considered before~\cite{Maiti2011,chubukov_review} so we will be quick. There  are two classes of eigenfunctions with
$s$-wave symmetry: one can be expanded in series of $\cos {4 n \phi_i}$ and $\cos {4 n \theta_i}$  ($n=0,1,2...$), where $\phi_i$ and $\theta_i$ are angles along hole and electron pockets, respectively, the other contains a series of terms $\cos {4 n \phi_i}$ and $\pm \cos{(4n+2) \theta_j}$, where the overall sign changes between the two electron pockets. The representatives of the two classes are $\cos k_x \cos k_y$ and $(\cos k_x + cos k_y)$. Expanding each of the two near $(0,0)$ or near $(\pi,\pi)$ and projecting onto $C_4-$symmetric hole pockets, we obtain a series of $\cos {4 n\phi_i}$ terms.  Expanding the same two $s$-wave terms near $(0,\pi)$ and $(\pi,0)$, we obtain series of  $\cos {4 n \theta_i}$ terms for the first and $\pm \cos{(4n+2) \theta_j}$ for the second.

The LAHA takes the leading terms in each of the two subsets (i.e., a constant term and $\pm \cos 2 \theta_j$ term) and neglects subleading terms. Within this approximation, $s$-wave gaps on the hole pockets are angle independent and the ones on electron pockets are $\Delta_e \pm {\bar \Delta}_e \cos{2\theta}$. The LAHA can be easily extended to include the  $\cos{4 \phi_i}$ dependencies.

\begin{widetext}

 \begin{figure}[t!]
 \includegraphics[width=1.0\textwidth]{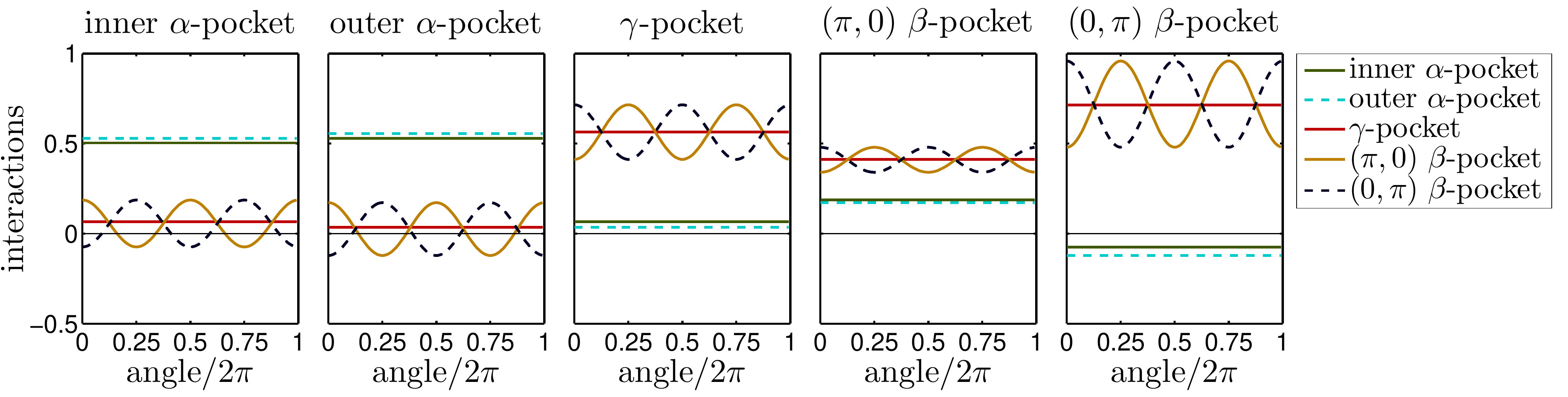}
     \caption{Behavior of the bare interactions $\Gamma^s_{i,j}({\bf k}_{F},{\bf k^\prime}_{F})$ between fermions on the Fermi surfaces $i$ and $j$ of LiFeAs, as obtained by LAHA procedure for $J=0.125U$. We set ${\bf k}_F$ to be along the $x$ direction on the FS labeled by $i$, and vary ${\bf k}_F^\prime$ along each of the FSs, labeled by $j$ ($i,j = 1-5$). The angle is measured relative to $k_x$.}
     \label{interaction_swave}
 \end{figure}

\end{widetext}

In explicit form, the $s$-wave components of the pairing interactions between different FSs are given by
\begin{align*}
\Gamma^s_{h_ih_j}(\phi,\phi^\prime)&=U_{h_ih_j}\\
\Gamma^s_{h_ie_{1},h_i e_2}(\phi,\theta)&=U_{h_ie}\left(1\pm2\alpha_{h_ie}\cos{2\theta}\right)\\
\Gamma^s_{e_1e_1,e_2e_2}(\theta,\theta^\prime)&=U_{ee}(1\pm2\alpha_{ee}(\cos{2\theta}+\cos{2\theta^\prime})\\
&+4\beta_{ee}\cos{2\theta}\cos{2\theta^\prime}\\
\Gamma^s_{e_1e_2,e_2e_1}(\theta,\theta^\prime)&=U_{ee}(1\pm2\alpha_{ee}(\cos{2\theta}-\cos{2\theta^\prime})\\
&-4\beta_{ee}\cos{2\theta}\cos{2\theta^\prime}\\
\end{align*}
where $h_1$ and $h_2$ correspond to the two $\alpha$ pockets, $h_3$ is $\gamma$ pocket, $e_{1,2}$ are the two electron pockets. For convenience of presentation we label the angles along hole pockets as  $\phi$ and $\phi'$ and the angles along electron pockets as $\theta$ and $\theta^\prime$.
In terms with $\Gamma^s_{ab,cd}$ the plus sign in the r.h.s. is for $ab$ combination and minus sign is $cd$ combination.
The interaction $U_{ee}$ is the sum of the interaction within a given electron pocket and inter-pocket interaction between the two electron pockets~\cite{Maiti2011}.
The coefficients -- the overall factors $U_{i,j}$ and the angle-dependent factors $\alpha_{i,j}$ are obtained from the orbital model using the procedure which we described above.  The values of  $U_{i,j}$ and $\alpha_{i,j}$ are obviously different for different $k_z$. In Tables I-II we show the results for three different values of $J$ (including the one used in Ref.\cite{Maiti2011}). In Fig.\ref{interaction_swave} we plot the interactions $\Gamma^s_{i,j}({\bf k}_{F},{\bf k^\prime}_{F})$ between fermions on the Fermi surfaces $i$ and $j$ for $J=0.125U$.
\begin{widetext}
\begin{center}
\begin{table}[hbtp]
\begin{center}
\tabcolsep=0.10cm
\begin{tabular}{cccccccccccccccc}
s-wave & $U_{h_1h_1}$ & $U_{h_2h_2}$ & $U_{h_3h_3}$ & $U_{h_1h_2}$ & $U_{h_1h_3}$ & $U_{h_2h_3}$ & $U_{h_1e}$ & $\alpha_{h_1e}$ & $U_{h_2e}$ & $\alpha_{h_2e}$ & $U_{h_3e}$ & $\alpha_{h_3e}$ & $U_{ee}$ & $\alpha_{ee}$ & $\beta_{ee}$ \\
\hline
$J=0.0U$&	 0.46	&	 0.50	&	 0.60	&	 0.47	&	 0.15	&	 0.12	&   0.14	&	 0.34	&	 0.22	&	 0.49	&	 0.60	&	 -0.12	&	 0.60	 &	 -0.12	&	 0.03\\
$J=0.125U$&	 0.50	&	 0.56	&	 0.56	&	 0.53	&	 0.07	&	 0.04	&   0.06	&	 1.16	&	 0.03	&	 2.90	&	 0.56	&	 -0.13	&	 0.56	 &	 -0.14	&	 0.04\\
$J=0.25U$&	 0.55	&	 0.62	&	 0.52	&	 0.58	&	 -0.01	&	 -0.05	&	 -0.03	&	 -3.33	&	 -0.06	&	 -1.60	&	 0.53	&	 -0.15	&	 0.53	 &	   -0.16	&	 0.05\\
\hline
\end{tabular}\caption{LAHA projected interactions in the $s-$wave channel for  \mbox{k$_z =\pi$}. The energies are in units of $U$.}
\end{center}
\begin{center}
\tabcolsep=0.10cm
\begin{tabular}{ccccccc}
s-wave & $U_{h_3h_3}$ & $U_{h_3e}$ & $\alpha_{h_3e}$ & $U_{ee}$ & $\alpha_{ee}$ & $\beta_{ee}$ \\
\hline
$J=0.0U$&	 0.76	&	 0.68	&	 -0.14	&	 0.61	&	 -0.13	&	 0.04\\
$J=0.125U$&	 0.74	&	 0.64	&	 -0.16	&	 0.57	&	 -0.15	&	 0.05\\
$J=0.25U$&	 0.71	&	 0.61	&	 -0.19	&	 0.53	&	 -0.17	&	 0.06\\
\hline
\end{tabular}\caption{LAHA projected interactions in the $s-$wave channel for \mbox{k$_z =0$}. The energies are in units of $U$.}
\end{center}
\label{table2}
\end{table}
\end{center}
\end{widetext}
We clearly see that the interactions between fermions within  subset I (two $\alpha$ pockets $h_1$ and $h_2$) and within subset II ($\gamma$ and $\beta$ pockets, $h_3$ and $e_{1,2}$) are far  stronger than the interactions between the two subsets. This is expected given the difference in the orbital content of FSs from subsets I and II.
We extended the analysis to include the $\cos{4 \phi}$  dependence of the interaction on the $\gamma$ pocket. We found, however, that the $\cos 4 \theta$
component is very weak: for $J = 0.125U$ we found, at $k_z = 0$,  $U_{h_3h_3} (\phi, \phi') = U_{h_3h_3} \left[ 1  -0.08 \left(\cos{4\phi} + \cos{4\phi'}\right) +0.02\cos{4\phi}\cos{4\phi'}\right]$, where $U_{h_3h_3}$ is the same as in Table \ref{table2}. Such a small interaction cannot give rise to any sizable $\cos{4\phi}$ dependence of the gap on the $\gamma$ pocket.  It is then likely that the observed $\cos {4\phi}$ dependence of the gap on the $\gamma$ pocket (Ref. \cite{Symmetry}) is due to the non-circular form of this FS, as an earlier study suggested~\cite{maiti_korsh}.

We now use the interactions  from Tables I and II as inputs, solve the  coupled linearized gap equations, and obtain the eigenvalues and the corresponding eigenfunctions (the gaps $\Delta_i$) in the $s$-wave channel.

\section{The structure of the {\it s}-wave superconducting gap}
\label{sec:4_sc_gap}

The  gaps on the hole and the electron pockets, consistent with $s-$wave interactions, are (for circular FSs)
\begin{align}
\begin{split}
\Delta_{h_1}(\phi)&=\Delta_{h_1}	\\
\Delta_{h_2}(\phi)&=\Delta_{h_2}	\\
\Delta_{h_3}(\phi)&=\Delta_{h_3}	\\
\Delta_{e_1}(\theta)&=\Delta_{e}+\bar{\Delta}_{e}\cos{2\theta}	\\
\Delta_{e_2}(\theta)&=\Delta_{e}-\bar{\Delta}_{e}\cos{2\theta}	\\
\end{split}\label{eq:swaveansatz}
\end{align}

We consider two characteristic $k_z$ cuts, one at $k_z=\pi$ and the other at $k_z=0$. For $k_z=\pi$, we solve the set of five coupled linearized gap equations, for $k_z = 0$ we solve a $3 \times 3$ set for $\Delta_{h_3}$, $\Delta_{e}$, and $\bar{\Delta}_{e}$. In each of the two cases we analyze the eigenfunctions with the largest and next to largest eigenvalues $\lambda_j$ ($j$ numbers the solutions). The eigenfunction with the largest $\lambda_j$  yields the best solution for the gap for a given $J/U$ in our one-parameter model. We gauge the relevance of the  solution with next-to-largest $\lambda$ by looking how close it is to the largest $\lambda$. If the two are close, the solution with next-to-largest $\lambda$ is a competitor, and may win upon a modest modification of the Hamiltonian (e.g., by lifting a restriction $J' = J$ and $U' = U - 5 J/2$ or including the renormalizations of the vertices by high-energy fermions).  If the largest and next-to-largest $\lambda's$  are substantially different, the best solution we found is unlikely to change under a modest variation of parameters. We will see below that, for renormalized interactions, the next-to-largest $\lambda$ is substantially smaller than the largest one, {\it i.e.}, it is sufficient to consider only the leading solution.  We will also see that for un-renormalized interactions $\lambda$ for the leading solution is very small, {\it i.e} $s$-wave superconductivity is "on the verge". The advantage of LAHA is that it allows us to identify how different interactions affect $s-$wave superconductivity and to understand the origin of the gap structure for the leading solution. In addition, it allows to study why the leading $\lambda$ is small, and what changes in the system parameters make $\lambda$ positive.

We now discuss the results. We first consider the gap structure for un-renormalized interactions and then analyze how the gap changes when we vary the strength of intra-pocket and inter-pocket interactions.

\subsection{s-wave gap for unrenormalized interactions}

\begin{widetext}

 \begin{figure}[t!]
 \includegraphics[width=1.0\textwidth]{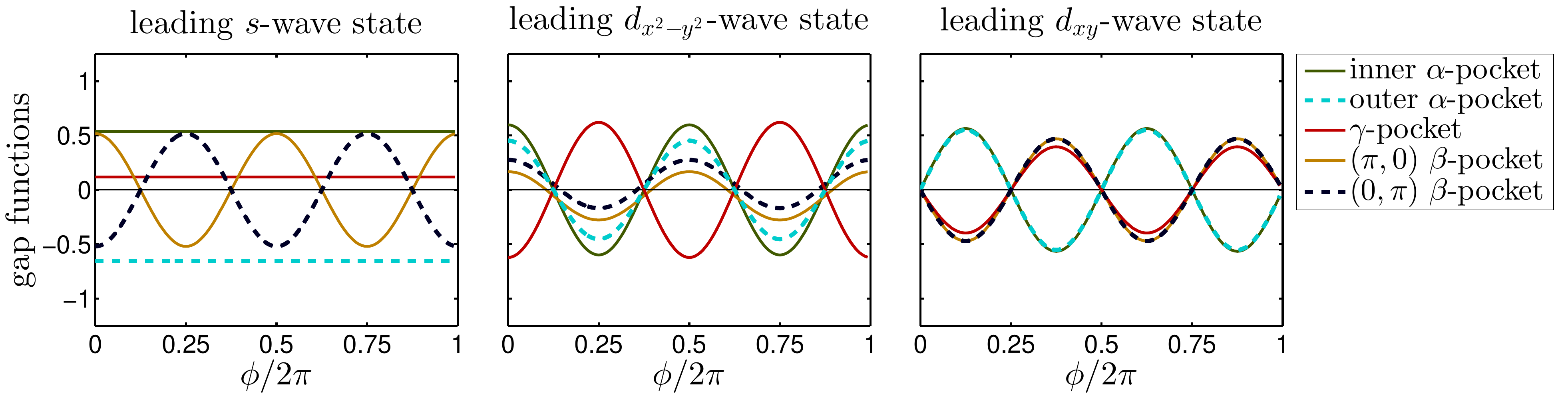}
     \caption{$s$-wave (left panel), $d_{x^2-y^2}$-wave (central panel), and $d_{xy}$-wave (right panel) leading eigenvalue solutions obtained by applying LAHA for the bare parameters and $J=0125U$.}
     \label{gaps_bare_swave}
 \end{figure}

\end{widetext}

At $k_z=0$, the solution of the $3\times3$ set yields a conventional $s^{+-}$ superconductivity with the sign change of the gap between hole and electron pockets (see Fig.\ref{fig:0})
and with $\pm \cos 2 \phi$ gap variation along electron pockets. We present the results in Table III and in the left panel of Fig.\ref{gaps_bare_swave}.

\begin{table}[hbtp]
\begin{minipage}[h]{0.33\textwidth}
\begin{center}
\tabcolsep=0.10cm
\begin{tabular}{cccccc}
					&\multicolumn{2}{c}{$k_z=\pi$}&	&\multicolumn{2}{c}{$k_z=0$}\\
\hline					
$\Delta_{h_1}$		&$+0.54$	&$+0.50$	&		&$$&$$\\
$\Delta_{h_2}$		&$-0.65$	&$-0.25$	&		&$$&$$\\	
$\Delta_{h_3}$		&$+0.12$	&$-0.34$	&		&$+0.84$	&$-0.14$\\	
$\Delta_{e}$		&$-0.00$	&$+0.05$	&		&$-0.43$	&$+0.23$\\	
$\bar{\Delta}_{e}$	&$+0.52$	&$-0.76$	&		&$+0.33$	&$+0.96$\\
\hline
$\lambda$			&$0.00$		&$-0.00$	&		&$-0.00$	&$-0.05$\\
\hline
\end{tabular}\caption{The two largest eigenvalues of the s-wave solution for $J=0.125U$ at the two different $k_z$ values. The overall scale for $\Delta_i$ is an arbitrary number. $\pm 0.00$ means that eigenvalue is positive (negative), but is smaller by magnitude than $5 * 10^{-3}$. For $k_z = \pi$, the two largest eigenvalues correspond to state A in
Fig. \ref{fig:1}.}
\end{center}
\end{minipage}
\end{table}

At  $k_z=\pi$ we found that the solution with the largest eigenvalue is the one which involves the sign change between the two tiny $\alpha$ hole pockets and between the larger hole pocket and the electron pockets (see Table III).  This is state A in Fig. \ref{fig:1}. In such a state, the gap on one of the $\alpha$ pockets is opposite to that on the $\gamma$ pocket. In our model, this sign change occurs between  the gap on  the $\gamma$ pocket and the outer $\alpha$ pocket, on which the gap has been measured in ARPES  experiments~\cite{Symmetry,umezawa}. (ARPES data obtained by the Dresden group indicates that  the inner hole pocket in LiFeAs may actually be buried under the FS~\cite{Symmetry}.) The exact values of the largest $\lambda$ are of order $10^{-13}$, i.e., are zero for all practical purposes.

For comparison, in Table IV we show the results for the gaps and the eigenvalues $\lambda_I$ and $\lambda_{II}$ for the case when we artificially cut the %
\begin{table}[hbtp]
\begin{center}
\tabcolsep=0.10cm
\begin{tabular}{cccccccc}
                                        &\multicolumn{4}{c}{$k_z=\pi$}&        &\multicolumn{2}{c}{$k_z=0$}\\
\hline
$\Delta_{h_1}$                &$     $        &$+0.72$        &$     $        &$+0.69$&|                        &$$&$$\\
$\Delta_{h_2}$                &$     $        &$-0.69$        &$     $        &$+0.72$&|                        &$$&$$\\
$\Delta_{h_3}$                &$+0.89$        &$     $        &$+0.14$        &$     $&|                        &$+0.84$        &$+0.14$\\
$\Delta_{e}$                &$-0.45$        &$     $        &$+0.07$        &$     $&|                        &$-0.43$        &$-0.23$\\
$\bar{\Delta}_{e}$        &$-0.06$        &$     $        &$+0.99$        &$     $&|                        &$+0.33$
&$-0.96$\\
\hline
$\lambda$                        &$-0.00$        &$-0.00$        &$-0.04$        &$-1.06$&|                        &$-0.00$        &$-0.05$\\
\hline
\end{tabular}\caption{s-wave solution for $J=0.125U$ when the coupling
between subset I and II is set to zero.}
\end{center}
\end{table}
interaction between the two subsets.  We see that the gap structure does not change much, but $\lambda_I$ and $\lambda_{II}$ are small
negative numbers, i.e., the actual $\lambda$ increases when the two subsets weakly interact with each other. The difference is minimal, though, as all eigenvalues are very close to zero.  We will see below that $\lambda \approx 0$ is the artefact of keeping the bare values of the interactions. Once we include modest modifications of the couplings, the largest $\lambda$ becomes of order one.

\subsubsection{Analytical reasoning}

These results can be understood analytically. Suppose first that the subsets I and II are decoupled. A set of linear gap equations can then be solved independently within each subset. For subset I the leading eigenvalue $\lambda_I$ is~\cite{maiti_11}
  \begin{align}
 \lambda_I = &-\frac{U_{h_1h_1} + U_{h_2h_2}}{2} + \nonumber \\
 & \sqrt{\left(\frac{U_{h_1h_1} - U_{h_2h_2}}{2}\right)^2 + U_{h_1h_2}^2}
 \label{eq:subsetA}.
\end{align}
It is positive when $U_{h_1h_2}^2 > U_{h_1h_1} U_{h_2h_2}$, i.e., when  inter-pocket interaction exceeds geometric mean of intra-pocket repulsions.
Because intra-pocket interaction is repulsive, the  corresponding eigenvector (proportional to the supercondicting order parameter) changes sign between the $\alpha$-pockets.  For our model parameters $U_{h_1h_1}U_{h_2h_2} \approx U_{h_1h_2}^2$ for all values of $J/U$ (see Table I). As a result,
\begin{align}
\lambda_I \approx \frac{U_{h_1h_2}^2 - U_{h_1h_1}U_{h_2h_2}}{U_{h_1h_1} + U_{h_2h_2}}
\label{eq_ex1}
\end{align}
is close to zero despite that $U_{h_1h_2}$, $U_{h_1h_1}$, and $U_{h_2h_2}$ are not small. Using the parameters from Table 1 we obtain $\lambda_{I} = -0.001$ for all three values $J/U =0, 0.125$, and $0.25$.

In the subset II, superconductivity is due to interaction between hole and electron pockets, like in many other FeSCs, and the susperconducting state is
 a conventional $s^{+-}$. If the interactions were angle-independent, $\lambda_{II}$ would be given by
\begin{align}
\lambda_{II} = -\frac{U_{h_3h_3} + 2 U_{ee}}{2} +  \sqrt{\left(\frac{U_{h_3h_3} - 2 U_{ee}}{2}\right)^2 + 2 U_{h_3e}^2}
\label{eq:subsetB}.
\end{align}
where, we remind, $U_{h_3 e}$ is intra-pocket hole-electron interaction  and  $U_{h_3 h_3}$ and $U_{ee}$ are
 the interaction within the $\gamma$ and the $\beta$ pockets, respectively.
 The condition $\lambda_{II} >0$ is $U_{h_3h_3}U_{ee}<U_{h_3e}^2$.
 In our model, $U_{h_3h_3}, U_{ee}$ and $U_{h_3e}$ are about the same  for $k_z = \pi$  (see Table I) and are different but satisfy $U_{h_3h_3}U_{ee} \approx U_{h_3e}^2$ for $k_z =0$ (see Table II).

For the actual case of angle-dependent interaction between $\gamma$ and $\beta$ pockets and between the two $\beta$ pockets, $\lambda_{II}$  is determined by the interplay between the difference of $U_{h_3h_3}U_{ee}$ and $U_{h_3e}^2$ and angle-dependent components of the interaction. The value of $\lambda_{II}$ is given by the solution of
\begin{widetext}
\begin{eqnarray}
 &&\lambda^3 + \lambda^2 \left(U_{h_3h_3} + 2 U_{ee}(1 + 4 \beta_{ee})\right) + 2\lambda\left(U_{h_3h_3}U_{ee}- U_{h_3e}^2 + 4U^2_{ee} (\beta_{ee} - \alpha^2_{ee}) + 2 \beta_{ee} U_{h_3h_3}U_{ee} - 2 \alpha^2_{h_3e} U_{h_3e}^2\right) \nonumber \\
&& - 8 U_{ee} \left((\alpha^2_{ee} -\beta_{ee})U_{h_3h_3}U_{ee} + U_{h_3e}^2 (\beta_{ee} + \alpha^2_{h_3 e} -2 \alpha_{ee} \alpha_{h_3e})\right) =0
\label{eq:ex2}.
\end{eqnarray}
\end{widetext}

Solving this equation for  our parameters, we obtain for $k_z =\pi$, $\lambda_{II} = -0.00040, -0.00037, -0.00033$ for $J/U =0, 0.125$, and $0.25$, respectively, and for $k_z=0$, $\lambda_{II} = -0.00006$ for all three $J/U$.
\begin{widetext}

 \begin{figure}[t!]
 \includegraphics[width=0.8\textwidth]{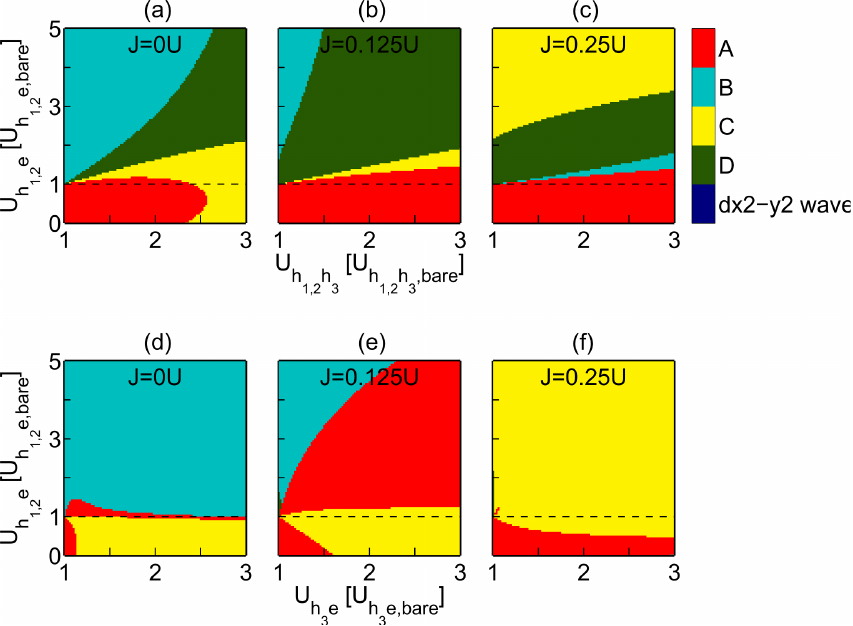}
 \caption{The phase diagram obtained by identifying the superconducting $s-$wave gap structure for given parameters
 with the eienfunction corresponding to the largest eigenvalue of the linearized gap equation. We vary the coupling either between two subsets I and II (the upper panel) or between electron and hole pockets (the lower panel). The bare (un-renormalized) values of parameters are the ones
 shown in Tables II.}
 \label{fig_new2}
 \end{figure}

 \end{widetext}
Let us now include a weak coupling between the two subsets.
For analytical understanding, we replace the actual model with five different FSs and momentum-dependent interactions by  a more tractable  3-gap model in which we assume that $s^{+-}$ superconductivity within the subset II is only weakly affected by the coupling between the subsets and approximate the subset II by a single eigenfunction (the gap) $\Delta_{II}$ and the corresponding eigenvalue $\lambda_{II}$, and  assume that sublets I and II are coupled  by an effective momentum-independent interaction $U_c$. In this situation, the eigenvalue $\lambda$ for the full system is obtained
by solving the $3\times 3$ set of linearized gap equations
\begin{eqnarray}
&&\Delta_{1} (U_{11}+\lambda) + U_{12} \Delta_2  + U_{c} \Delta_{II} =0,\quad \nonumber \\%
&&\Delta_{2} (U_{12}+ \lambda)  + U_{22} \Delta_2  +U_{c} \Delta_{II}  =0 ,\quad  \nonumber \\
&&\Delta_{II} (\lambda-\lambda_{II}) + U_c (\Delta_1 + \Delta_2) =0.
\label{eq:n3}
\end{eqnarray}
where $U_{11} \equiv U_{h_1 h_1}, U_{22} \equiv U_{h_2 h_2}, U_{12} \equiv U_{h_1 h_2}$.
In these notations,
$\lambda_I = - (U_{11} + U_{22})/2 + \left((U_{11} - U_{22})^2/4 + U^2_{12}\right)^{1/2}$.

The solution for $\lambda$ can easily be obtained by solving the cubic equation on $\lambda$.
To get a feeling for the effect of a small inter-subset coupling $U_c$, let's take $\lambda_I$ and $\lambda_{II}$ to be zero, {\it i.e.},
set $U_{11}U_{22} =U_{12}^2$.  One can easily make sure that for a non-zero $U_c$, the set (\ref{eq:n3}) has one positive solution. To first order in $U_c$, a positive $\lambda$ is
 \begin{equation}
\lambda \approx |U_c| \left(\frac{(\sqrt{U_{11}} - \sqrt{U_{22}})^2}{U_{11} + U_{22}}\right)
\label{eq:n8}
\end{equation}
Although $\lambda$ is positive, it is rather small because $U_{11}$ and $U_{22}$ are quite close. This is in line with what we obtained numerically.
The ratio of the gaps on the $\alpha$ pockets is
\begin{equation}
\left|\frac{\Delta_1 -\Delta_2}{\Delta_1 + \Delta_2}\right| \approx \frac{\sqrt{U_{11}} + \sqrt{U_{22}}} {\sqrt{U_{11}}-\sqrt{U_{22}}}
\label{eq:n11}
\end{equation}
For $U_{11} \sim U_{22}$,  we have $\Delta_1 \approx -\Delta_2$ i,e., the superconducting state is the A state.
We see therefore that, when $\lambda_{I}$ and $\lambda_{II}$ are near zero, as in our case, the coupling between the subsets keeps the $s^{+-}$ gap structure within each subset, but increases the value of the eigenvalue $\lambda$. In particular, as we just found, $\lambda$ immediately  becomes positive at a non-zero $U_c$ if $\lambda_I = \lambda_{II} =0$.  This agrees with the numerical results in Table III.  The relative phase between the two subsets (e.g., the sign of the gap on the $\gamma$ pocket compared to that on the inner $\alpha$ pocket) is determined by the sign of $U_c$. We see from Table III that the two leading eigenstates can be modeled by coupling the subsets I and II by $U_c$ of different signs.

\subsection{s-wave gap for renormalized interactions}

We now consider what happens if we vary the interactions within each subset and between the two subsets.
There is no clear experimental evidence that in LiFeAs the renormalizations in one particular channel are stronger than in the other
(e.g., there is no evidence that the renormalizations in the particle-hole spin channel play the dominant role), and we check what happens when we vary different interaction components.

\begin{widetext}

\begin{figure}[t!]
\includegraphics[width=0.8\textwidth]{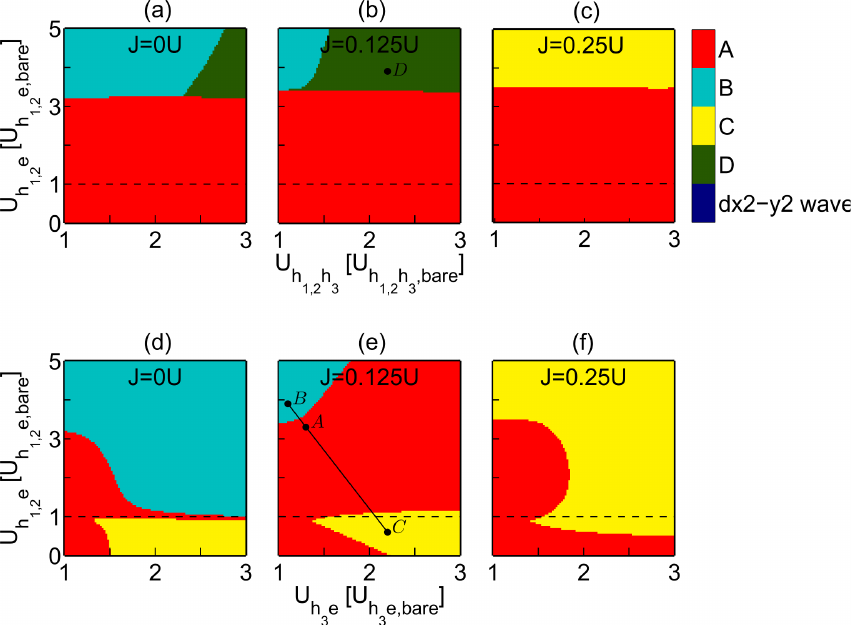}
\caption{
 The same as in Fig. \ref{fig_new2} but with $U_{h_1h_2}$ increased by a factor 1.5. The solid line BAC in (e) is the direction chosen in  Figs.\ref{fig3},\ref{fig4}. The points A, B, C, D refer to the values of the interaction, employed to obtain the superconducting gaps plotted in Fig.\ref{fig:gaps}.}
\label{fig_new2_1}
\end{figure}

\end{widetext}
We vary the interactions in two different ways.  First we increase/decrease the interactions between the two subsets, {\it i.e.}, vary $U_{h{1,2}h_3}$ and $U_{h_{1,2}e}$. We show the results in the upper panel of Fig. \ref{fig_new2}. We recall that for un-renormalized interactions we obtained the A state, however the eigenvalue $\lambda$ is close to zero. We see from Fig. \ref{fig_new2} that under the change of the interactions the system either remains in the state A (and $\lambda$ increases, as we show below), or transform into one of three other states --states B, C, or D, which we presented schematically in Fig. \ref{fig:1}. Specifically,  the state A survives when we predominantly increase $U_{h_{1,2}h_3}$, the state B (a conventional $s^{+-}$ state) wins at small $J/U$  when we predominantly increase $U_{h_{1,2}e}$,  the state C wins at larger $J/U$, also when we predominantly increase $U_{h_{1,2}e}$, and the state $D$ wins when we increase $U_{h_{1,2}h_3}$ and $U_{h_{1,2}e}$ by about the same amount. Note that all four states can be realized by modestly varying the interactions.
This is obviously the consequence of the fact that for bare interactions we are dealing with a near-critical situation of two weakly coupled subsets, each with
almost vanishing eigenvalue.

The change of the s-wave gap structure is expected as, e,g., when we increase  $U_{h_{1,2}e}$ at small $J/U$, we increase a repulsive interaction
between fermions at $\alpha$ and $\beta$ pockets, and, when this repulsion is strong enough, the system prefers the state B from Fig. \ref{fig:1}, in which the sign of the gaps on both $\alpha$ pockets is opposite to that on the $\beta$ pockets. At larger $J/U=0.25$, $U_{h_{1,2}e}$ is attractive, and, when this interaction gets larger, the system prefers the state C from the same Figure, in which the sign of the gap on  the $\alpha$ and $\beta$ pockets is the same.
The state D is realized when there is a strong repulsion between $\alpha$ pockets and both $\gamma$  and $\beta$ pockets. Then the system prefers the state in which the gap on the $\alpha$ pockets is of one sign, and the gap on $\beta$ and $\gamma$ pockets is of the opposite sign.

In the lower panel in Fig.  \ref{fig_new2} we show the phase diagram for the case when we vary interactions in a different way -- by changing the interactions between hole and electron pockets.  An increase of $\alpha-\beta$ and $\gamma-\beta$ interactions is expected if spin-fluctuations are strong an peaked at momentum transfers $(0,\pi)$ and $(\pi,0)$ in the unfolded 1Fe zone ($(\pi,\pi)$ in the folded 2Fe zone). For repulsive interactions between $\alpha$ and $\beta$ pockets ($J=0$ and $J =0.125U$), we found the B phase in a wider range. This is entirely expected as the B phase is the anticipated  result when spin fluctuations are strong. For attractive interaction between $\alpha$ and $\beta$ pockets ($J =0.25U$), we found  a wider range of the C phase.  Again, this is an expected result  because, as we said, strong attractive  $U_{h_{1,2}e}$ favors  the state in which the sign of the gap on  the $\alpha$ and $\beta$ pockets is the same.   At the same time, strong repulsion $U_{h_{3}e}$ forces the gaps on $\gamma$ and $\beta$ pockets to be of the opposite sign.  Note that the phase D essentially disappears and gets replaced by the phase C.

In Fig. \ref{fig_new2_1} we show  the results for the same variation of system parameters as in Fig. \ref{fig_new2}, but, in addition, we increase $U_{h_1h_2}$ by a factor 1.5. This is done to model the case when there is an additional increase of the interaction at small momentum transfer. We see all four phases, as in the previous figure, but now there is much wider region of the A phase, particularly in the upper panel. This is again an expected result as larger repulsion between the two $\alpha$ pockets  stabilizes the state with plus-minus gap between the $\alpha$ pockets, which is our A state.

In Table V we show the solution for the gaps and the eigenvalues for the leading and subleading solutions for the representative parameter set when $J =0.125 U$, $U_{h_{3}e}$ and  $U_{h_1h_2}$  are increased by  1.5, and $U_{h_{1,2}e}$  is increased by 1.75 compared to the values in Tables I and II. For $k_z =\pi$ this corresponds to the point A in Fig. \ref{fig_new2_1} e.  We see that the state A has the largest eigenvalue and much larger $\lambda$ than for bare parameters.  This obviously leads to a much larger $T_c$ than for bare parameters.

\begin{table}[hbtp]
\begin{minipage}[h]{0.33\textwidth}
\begin{center}
\tabcolsep=0.10cm
\begin{tabular}{cccccc}
					&\multicolumn{2}{c}{$k_z=\pi$}&	&\multicolumn{2}{c}{$k_z=0$}\\
\hline							
$\Delta_{h_1}$		&$+0.65$	&$+0.07$	&		&$$&$$\\
$\Delta_{h_2}$		&$-0.52$	&$+0.48$	&		&$$&$$\\	
$\Delta_{h_3}$		&$+0.46$	&$+0.08$	&		&$+0.85$	&$+0.03$\\
$\Delta_{e}$		&$-0.30$	&$-0.17$	&		&$-0.48$	&$-0.17$\\	
$\bar{\Delta}_{e}$	&$-0.08$	&$-0.86$	&		&$+0.19$	&$-0.98$\\
\hline
$\lambda$			&$+0.30$	&$+0.26$	&		&$+0.27$	&$-0.05$\\
\hline
\end{tabular}\caption{s-wave solution for $J=0.125U$ and for modified interactions:
 $U_{h_{3}e}$ and  $U_{h_1h_2}$  are increased by  1.5 and $U_{h_{1,2}e}$  is increased by 1.75 compared to the values in Tables I and II.
For $k_z =\pi$, this corresponds to point A in  Fig. \ref{fig_new2_1} e.}
\end{center}
\end{minipage}
\label{table8}
\end{table}

\begin{figure}[htbp]
\includegraphics[width=0.4\textwidth]{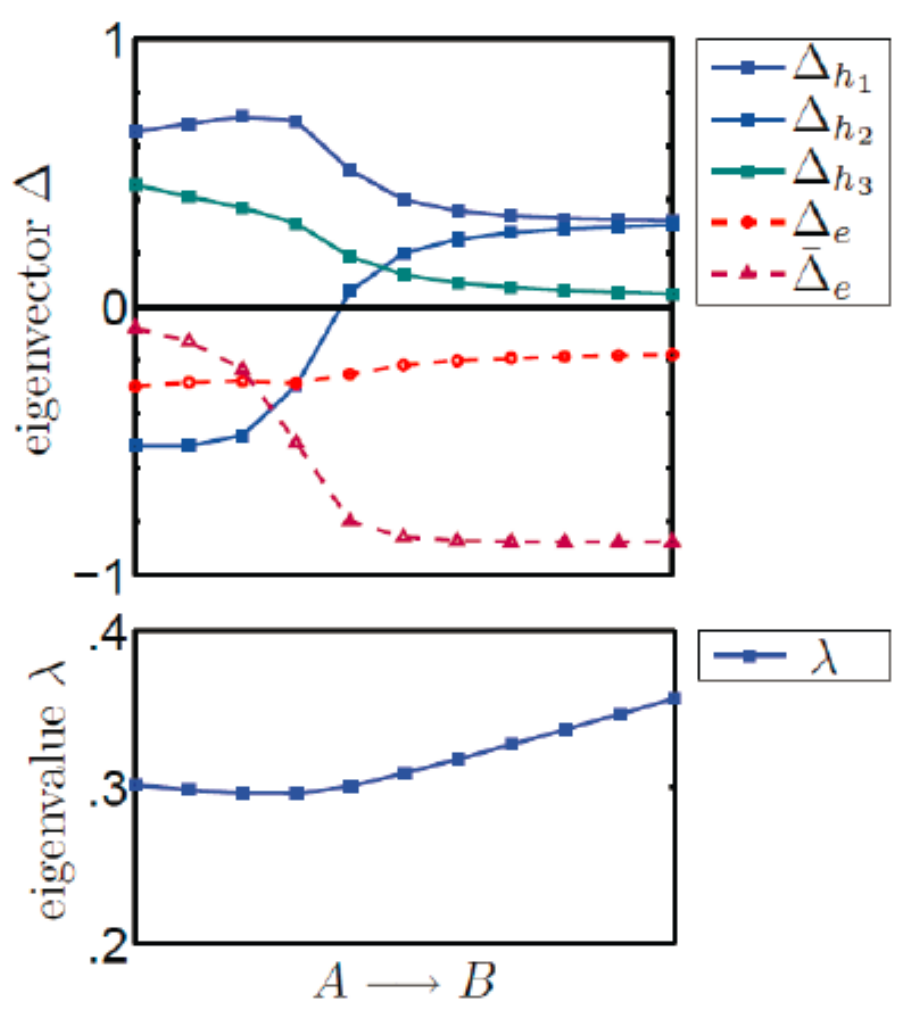}
\caption{Evolution of the gap structure (upper panel) and the largest eigenvalue (lower panel)  along AB line in Fig. \ref{fig_new2_1} e.
The two limiting states are A and B from Fig. \ref{fig:1}.}
\label{fig3}
 \end{figure}
\begin{figure}[htbp]
\includegraphics[width=0.4\textwidth]{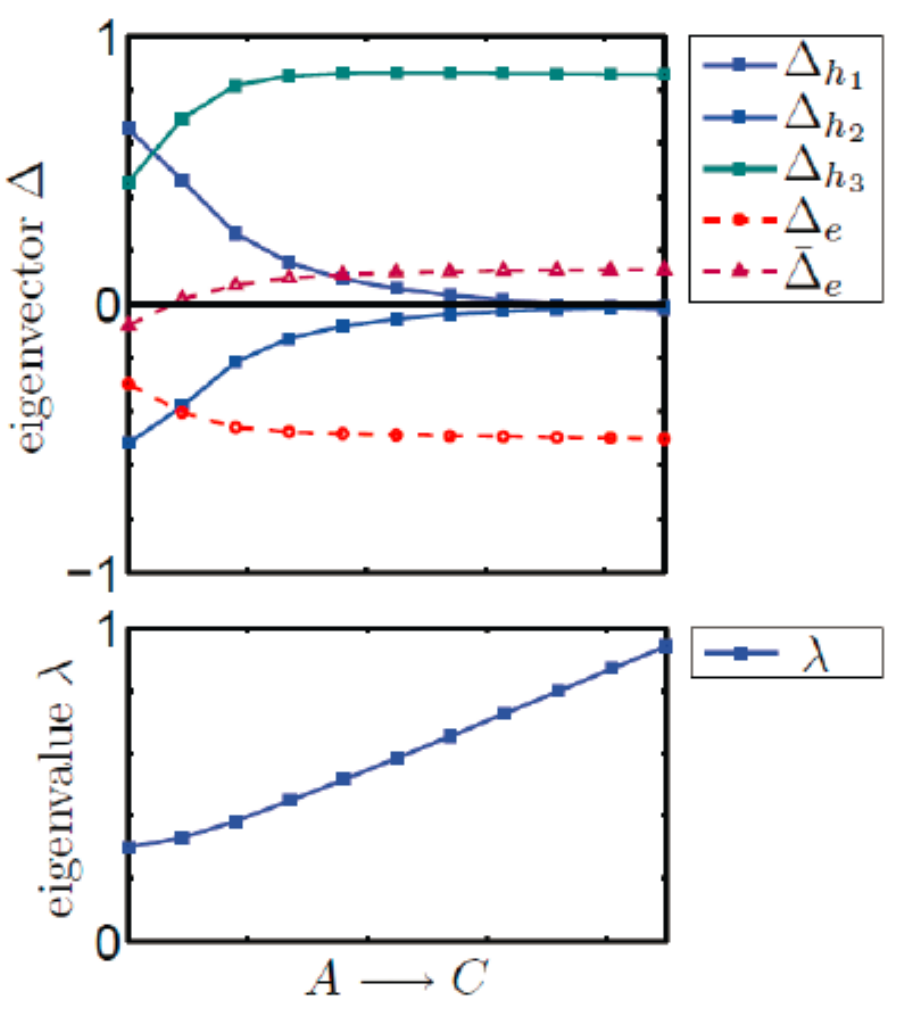}
\caption{Same as in Fig. \ref{fig3} but along AC line in
Fig.\ref{fig_new2_1}e. The two limiting states are A and C from Fig. \ref{fig:1}.}
\label{fig4}
\end{figure}

We also see from Table V that the state with second-to-largest $\lambda$ is the B state.  This is not surprising as the boundary of the B state is quite close to point A in  Fig. \ref{fig_new2_1} e.  In Fig. \ref{fig3} we show how the gap structure evolves as the system moves along AB line in Fig. \ref{fig_new2_1} e. We see that the evolution is  continuous. As the system moves along AB line, the gap on one of the $\alpha$ pockets shrinks, passes through zero, and then re-appears with the opposite sign.  The other gaps do evolve but preserve signs. It is essential for comparison with the ARPES data that the angle-dependent $\cos{2\theta}$  component of the gap on electron pocket (${\bar \Delta}_e$) increases as the system moves towards the B phase. We found that in most of the parameter range of the B phase in  Fig. \ref{fig_new2_1} e the gap has accidental nodes on electron pockets.

In Fig. \ref{fig4} we show the variation of the gaps and of the largest eigenvalue as the system moves in the opposite direction -- along AC line towards the C phase.  We see that the evolution is again continuous, but now the gap on the other $\alpha$ pocket gets smaller, passes through zero, and re-appears with a different sign. Interestingly, the gap on the other $\alpha$ pocket also gets smaller, although does not change sign.  As a result, inside the C phase, the gaps on the $\alpha$ pockets are much smaller than the gaps on $\gamma$ and $\beta$ pockets. The gap on the $\gamma$ pocket is the largest, consistent with the fact that this gap has a sign opposite to that for all other gaps. Note also that $\cos{2\theta}$ component of the gap on $\beta$ pockets changes sign but does not grow much as the system moves along AC line, i.e., the superconducting gap inside the C phase has no accidental nodes.

The evolution from the A state to the D state occurs in a similar way but involves the vanishing and eventual sign change of the gap on one of $\alpha$ pockets and on $\gamma$ pocket. The two sign changes generally happen at different values of parameters.

\subsubsection{Analytical reasoning}

The evolution from the state A to the other states can be traced within our analytical reasoning.  For the same model with inter-pocket interaction $U_c$, which we used in the previous section, the eigenvalue $\lambda$ evolves from the one in Eq. (\ref{eq:n8}) at small $U_c$ to
\begin{equation}
\lambda \approx \sqrt{2} |U_c|
\label{eq:n9}
\end{equation}
when $U_c$ becomes the largest interaction.  Simultaneously, the ratio of the gaps on the two $\alpha$ pockets becomes
\begin{equation}
\left|\frac{\Delta_1 -\Delta_2}{\Delta_1 + \Delta_2}\right| \approx \frac{U_{11} -U_{22}} {2 \sqrt{2} |U_c|}
\label{eq:n12}
\end{equation}
For $U_{11} \sim U_{22}$,  we have $\Delta_1 \approx \Delta_2$, {\it i.e.}, the gaps on the two $\alpha$ pockets
are of the same sign.  Whether the resulting state is state B or state C depends on the sign of $U_c$.

The analysis of the system evolution with increasing $U_c$ is straightforward, and the result is that the evolution is continuous:
as $|U_c|$  increases, the gap on one of the $\alpha$ pockets (either $\Delta_1$ or $\Delta_2$, depending on the sign of $U_c$) passes through zero and re-appears with a different sign. The implication of this result is that a transformation between the  states A and either B or C (or D)
is continuous along the $T_c$ line.

There is one limiting case, though, when the system evolution is not continuous.  This happens when intra-pocket repulsions $U_{11}$ and $U_{22}$ are identical. The solution of the set (\ref{eq:n3}) is particularly simple in this case: there is one negative solution (which describes $s^{++}$ superconductivity) and two  potentially positive solutions. These two are
\begin{eqnarray}
&&\lambda_1 = \lambda_I = - U_{11} + U_{12} \nonumber \\
&&\lambda_2  = \frac{\lambda_{II} -U_{11}-U_{12}}{2} + \sqrt{\frac{(\lambda_B +U_{11}+U_{12})^2}{4} + 2 U^2_c} \nonumber \\
&& \approx \lambda_{II} + \frac{2 U^2_c}{\lambda_B + U_{11} + U_{12}}
 \label{eq:n4}
\end{eqnarray}
where $\lambda_I$ and $\lambda_{II}$ are given by Eqs. (\ref{eq:subsetA}) and (\ref{eq:subsetB}), and we recall that
$U_{11} \equiv U_{h_1 h_1}, U_{22} \equiv U_{h_2 h_2}, U_{12} \equiv U_{h_1 h_2}$.
The eigenfunction corresponding to $\lambda_1=\lambda_I$ is the A-state, with zero gap for the subset $II$, the eigenfunction corresponding to $\lambda_2$  is either B or C state, depending on the sign of $U_c$.  If $\lambda_I > \Lambda_{II}$ and $\lambda_I >0$, the system orders into the state A small $U_c$, and then, at the critical $U_c$ given by
\begin{equation}
|U_c| \approx (\lambda_I -\lambda_{II})^{1/2}  \left(\frac{\lambda_{II} + U_{11} + U_{12}}{2}\right)^{1/2},
\label{eq:n7}
\end{equation}
undergoes a discontinuous transition into the state B or state C.  This behavior holds along the $T_c$ line in Fig. \ref{fig2}(a). If $\lambda_{II} > \lambda_I$,  the linearized gap equation yields only B state or C state. Note that these states emerges even if $\lambda_{II} <0$, because $\lambda_2$ is definitely positive for large enough $U_c$.

\subsection{The selection of the s-wave state -- the comparison with the APRES data}

Which of the four s-wave gap configurations (A,B,C, or D)  describes the superconducting state in  $LiFeAs$ depends on the strength of the dressed interactions in this material. If $(\pi,\pi)$ spin fluctuations  were strong enough, state B would be the most natural candidate. However, there is still no complete experimental evidence that spin fluctuations with large momentum transfer are strong\cite{knolle12}.
There is indirect evidence that spin-fluctuations with a small momentum transfer may also be present -- ferromagnetic fluctuations in the non-superconducting Li$_{1-y}$Fe$_{1+y}$As with $y \leq 0.04$ are quite strong~\cite{blundell13}.

In the absence of a guide from magnetic measurements, the only way to distinguish between the states A,B, C, and D is to compare the gap structure in each of these states with ARPES measurements and find the best match. We recall in this regard that in the state A the gap is the largest on the $\alpha$ pockets,
the gap on the $\gamma$ pocket is a bit larger than the FS-averaged gaps on the $\beta$ pockets, and there are no accidental nodes. In the state B, the gaps on the electron pockets are also the largest, but, at least in some range of parameters, there are strong $\cos 2 \theta$ gap variations on the electron pockets and accidental nodes (see Fig. \ref{fig3}). ARPES data on the other hand show quite convincingly that the gap on electron pockets does have $\cos 2 \theta$ component, but  not strong enough to cause accidental nodes\cite{Symmetry}. Still, for other parameters the nodes disappear, and we cannot rule out the state B as a potential superconducting state in LiFeAs. The states C  and D are less likely candidates as the state C has the largest gap on the $\gamma$ point, what is inconsistent with ARPES, and the state D has the nodes on electron pockets in the large portion of the parameter space where this state emerges. Overall, from the comparison with ARPES, the state A appears to be the most likely candidate.  If this is indeed the case, superconductivity in LiFeAs is unique in the sense that  the gap structure in this material is qualitatively different from the one in other Fe-based superconductors with hole and electron pockets.

We propose another experiment to  verify  whether  the gap structure in LiFeAs is the same as in the  state A. Namely, as we found, the state A has the highest eigenvalue if LeFeAs can be viewed as the system of weakly coupled subsets I and II, with near-zero eigenvalues $\lambda_I$ and $\lambda_{II}$ within each of the two subsets. The actual eigenvalue $\lambda$ gets larger due to the coupling between subsets I and II. This, however, occurs only near $k_z =\pi$, where both subsets are present, but not near $k_z=0$, where only subset II is present. As a consequence, if different regions of $k_z$ could become superconducting independent of each other, the region near $k_z =\pi$ would have a higher $T_c$. In real situation, the whole 3D material indeed becomes a superconductor at the same $T_c$, which is close to the one at $k_z =\pi$. Below this $T$, superconductivity at small $k_z$ is induced by proximity.
However, the fact that the coupling $\lambda$ is larger near $k_z =\pi$ than near $k_z = 0$ still implies that the temperature dependence of the gaps is different in the two regions. Near $k_z = \pi$, the gap(s) follow BCS behavior as functions of $T/T_c$, where $T_c$ is the actual superconducting temperature. Near $k_z =0$, the gaps remain small down to  a smaller $T < T_c$, at which subset II would become superconducting on its own, and increases only below this temperature.  It will be very interesting to verify this behavior by comparing the temperature dependencies of the gaps, extracted from ARPES measurements below $T_c$.

\section{System behavior at $T < T_c$.}
\label{below_tc}

\begin{widetext}

\begin{figure}[htbp]
\includegraphics[width=0.9\textwidth]{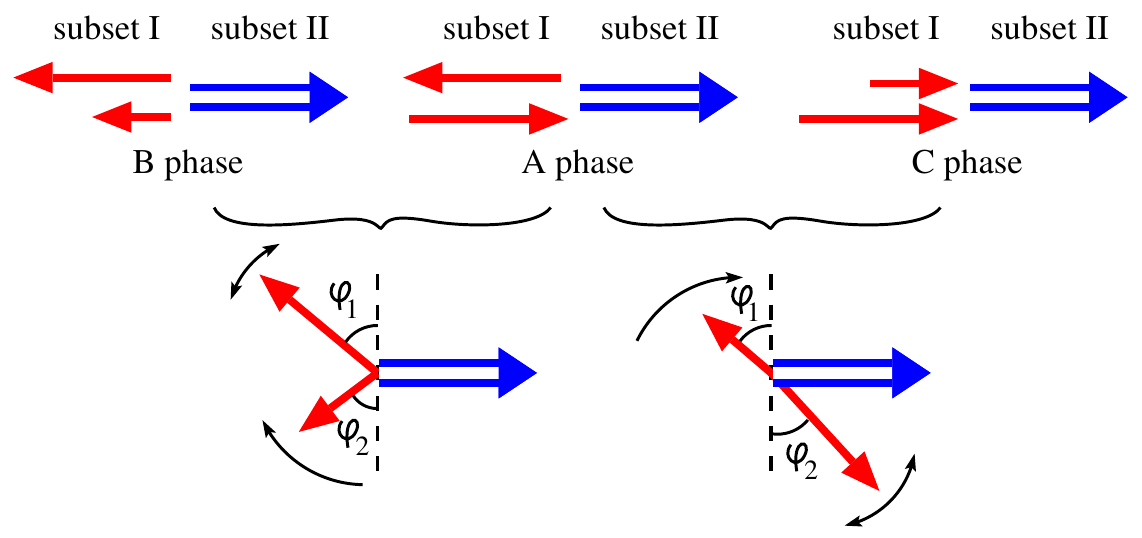}
\caption{An illustration  how the solutions which break time-reversal symmetry appear as the system parameters evolve along AB and AC lines in Fig. \ref{fig_new2_1}e   at $T < T_c$.   The double arrow represents the gap in the subset II.  The single lines represent the gaps on the two $\alpha$ pockets which constitute subset I.  The two intermediate states  can be termed as $A+iB$ and $A+iC$.}
\label{fig:E}
\end{figure}

\end{widetext}

So far we only considered the behavior immediately below $T_c$. A new behavior may emerge at a lower $T$, particularly in a situation when $\lambda_I$ and $\lambda_{II}$ are comparable in strength, as it is the case for LiFeAs. Namely, at low $T$, the evolution from state A to state B or state C is  is not necessary continuous even when $U_{11} \neq U_{22}$. Instead of changing the magnitudes of the gaps, as it happens at $T_c$, when the system moves along AB or AC lines, the system prefers to keep the magnitude of the gap fixed to maximize the condensation energy, and change instead the phase of the gap. This gives rise to an intermediate $s+is$ state  in which the phases of the gaps on $\alpha$ pockets differ by less than $\pi$, i.e., time-reversal symmetry is broken. We show these states schematically Fig. \ref{fig:E}. In our notations, such states can be termed as A+iB or A+iC.

\begin{widetext}

\begin{figure}[h!t]
\includegraphics[width=0.8\textwidth]{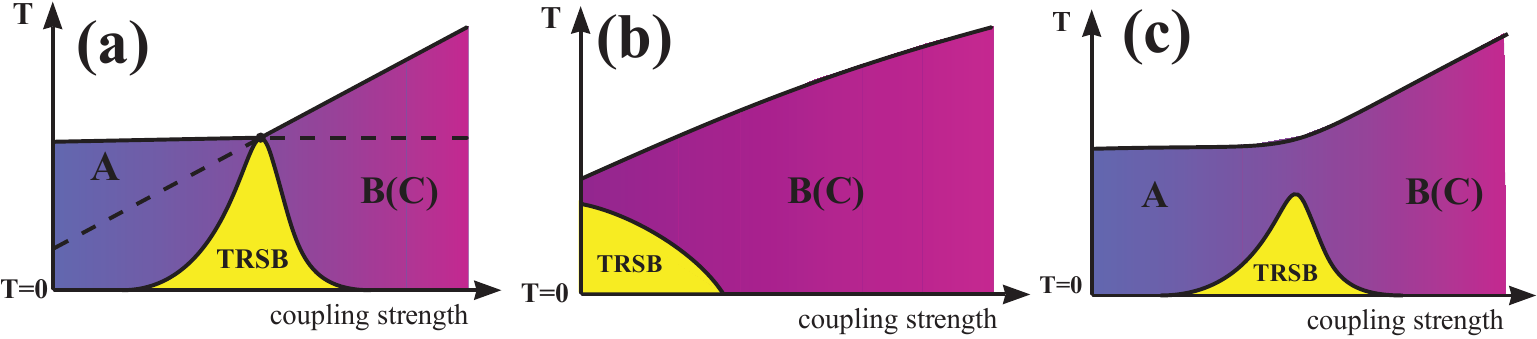}
\caption{The schematic phase diagrams for the system evolution with the change of the system parameters, e.g., along AB or AC lines in Fig. \ref{fig_new2_1}e.  TRSB stand for the intermediate phase in which superconductivity is s-wave, yet it breaks time-reversal symmetry. Depending on the system parameters, the intermediate phase either occurs for all temperatures below $T_c$ (panel a) or only at $T << T_c$.}
\label{fig2}
\end{figure}

\end{widetext}

Consider as an example the transition from the A to the B phase using our analytical reasoning within the effective three-band model with $\Delta_1, \Delta_2$, and $\Delta_{II}$. As we found above, at $T_c$,  the transformation from the  state A to the state B upon increasing $U_c$ is smooth, except when $U_{11} = U_{22}$. In the latter case, the transition from the pure  state A with zero gap in the subset II to the pure state B is discontinuous and occurs at a critical $U_c$, given by Eq. (\ref{eq:n7}).
The set of non-linear $3 \times 3$ equations has been analyzed in  Refs.\cite{Maiti2013,lara}, and we borrow the results of these works.
It turns out that, below $T_c$,  the system evolution with increasing $U_c$ may occur via an intermediate phase. In particular, when $U_{11} = U_{22}$ and $\lambda_I >0$ and larger than $\lambda_{II}$, the analysis shows~\cite{Maiti2013,lara} that there necessary exists  an intermediate state, which begins at $T_c$ at a critical $U_c$  and extends to a finite range of $U_c$ at $T < T_c$. In this intermediate state we have,  up to an overall phase factor, $\Delta_1 = \Delta e^{i\psi}$, $\Delta_2 = \Delta e^{-i\psi}$ and $\Delta_{II} = -a \Delta$, $a>0$. The two phases evolve between $\psi =0$ at the end of the intermediate state at a larger $U_c$ and $\psi = \pm \pi/2$ at the other end. In the intermediate phase, the value of $\psi$ can be either positive ($+\pi/2$ at the lower end) or negative ($-\pi/2$ at the lower end). The states with $\psi$ and $-\psi$ are related by time inversion, and the system spontaneously breaks time-reversal symmetry by selecting  either $+\psi$ or $-\psi$. Such a state has been labeled~\cite{Maiti2013,lara} as $s+is$. We show the  full phase diagram for this case in Fig. \ref{fig2}a.

When $\lambda_{II} > \lambda_{I}$, only the B  state or C state emerge at $T_c$. However, the calculations show that the intermediate $A +iB$ or $A + iC$  states still emerge at a smaller $T$, if the values of $\lambda_I$ and $\lambda_{II}$ are close. We show the phase diagram for this case in Fig. \ref{fig2}b.

When $U_{11} \neq U_{22}$, the behavior along $T_c$ line is smooth, no matter what the ratio $\lambda_I/\lambda_{II}$ is.  However, if  both $\lambda_I$ and $\lambda_{II}$ are small  and $s-$wave superconductivity is induced by the coupling between the subsets,  $s+is$ state still emerges below some $T<T_c$. We show the corresponding phase diagram in Fig.\ref{fig2}c.  It will be very interesting to analyze the behavior of LiFeAs below $T_c$ and verify whether there is some evidence for a second phase transition at $T < T_c$.

A question which we did not address in this work  is how the gap in the subset II evolves with $k_z$ in the parameter range where the state with the largest $\lambda$ at $k_z =\pi$ is the D state, with different gap structure in the subset II from the one at $k_z =0$.  It is quite possible that in this situation there exists another time-reversal symmetry breaking state with different phases of the gaps on the two electron pockets at intermediate $k_z$. This question is rather academic though, as the D state is an unlikely candidate for the superconducting state in LiFeAs.
\section{$d_{x^2-y^2}$ and $d_{xy}$ gaps structure}
\label{d-wave}

For completeness we also present the solution for superconductivity in the $d-$wave channel.
ARPES data do rule out $d-$wave pairing in LiFeAs, but it is still important to know whether one or both $d-$wave channels are attractive, and, if so, how close the $d-$wave eigenvalue is to the one in the $s-$wave channel, i.e., whether one can induce a $d-$wave instability by external perturbation, like pressure.

For $d_{x^2-y^2}$ channel, there are two subsets of angle-dependent terms and we take the leading eigenfunctions in both of them, like we did in  the $s-$wave channel. The result is
\begin{align*}
\Gamma^{d_{x^2-y^2}}_{h_ih_j}(\phi,\phi^\prime)&=\tilde{U}_{h_ih_j}\cos{2\phi}\cos{2\phi^\prime}\\
\Gamma^{d_{x^2-y^2}}_{h_ie_{1,2}}(\phi,\theta)&=\tilde{U}_{h_ie}\cos{2\phi}(\pm1+2\tilde{\alpha}_{h_ie}\cos{2\theta})\\
\Gamma^{d_{x^2-y^2}}_{e_1e_1,e_2e_2}(\theta,\theta^\prime)&=\tilde{U}_{ee}(1\pm2\tilde{\alpha}_{ee}(\cos{2\theta}+\cos{2\theta^\prime})\\
&+4\tilde{\beta}_{ee}\cos{2\theta}\cos{2\theta^\prime})\\
\Gamma^{d_{x^2-y^2}}_{e_1e_2,e_2e_1}(\theta,\theta^\prime)&=\tilde{U}_{ee}(-1\mp2\tilde{\alpha}_{ee}(\cos{2\theta}-\cos{2\theta^\prime})\\
&+4\tilde{\beta}_{ee}\cos{2\theta}\cos{2\theta^\prime})
\end{align*}
The couplings ${\tilde U}_{i,j}$ and the  prefactors ${\tilde \alpha}_{i,j}$ for the $\cos{2\theta}$ terms are obtained using the same procedure as for $s-$wave. The results are presented in Tables VI-VII and in Fig.\ref{fig:interactions_dx2y2}

\begin{widetext}
\begin{center}
\begin{table}[hbtp]
\begin{center}
\tabcolsep=0.10cm
\begin{tabular}{cccccccccccccccc}
$d_{x^2-y^2}$-wave & $\tilde{U}_{h_1h_1}$ & $\tilde{U}_{h_2h_2}$ & $\tilde{U}_{h_3h_3}$ & $\tilde{U}_{h_1h_2}$ & $\tilde{U}_{h_1h_3}$ & $\tilde{U}_{h_2h_3}$ & $\tilde{U}_{h_1e}$ & $\tilde{\alpha}_{h_1e}$ & $\tilde{U}_{h_2e}$ & $\tilde{\alpha}_{h_2e}$ & $\tilde{U}_{h_3e}$ & $\tilde{\alpha}_{h_3e}$ & $\tilde{U}_{ee}$ & $\tilde{\alpha}_{ee}$ & $\tilde{\beta}_{ee}$ \\
\hline
$J=0.0U$& 	 0.36	&	 0.39	&	 0.04	&	 -0.38	&	 0.12	&	 -0.13	&	 0.08	&	 0.71	&	 -0.08	&	 0.71	&	 0.03	&	 0.71	&	 0.02	 &	 0.71	&	 0.51\\
$J=0.125U$&  0.31	&	 0.34	&	 0.04	&	 -0.33	&	 0.11	&	 -0.11	&	 0.07	&	 0.71	&	 -0.07	&	 0.71	&	 0.02	&	 0.71	&	 0.01	 &	 0.71	&	 0.51\\
$J=0.25U$& 	 0.27	&	 0.29	&	 0.03	&	 -0.28	&	 0.09	&	 -0.10	&	 0.06	&	 0.71	&	 -0.06	&	 0.71	&	 0.02	&	 0.71	&	 0.01	 &	 0.71	&	 0.51\\
\hline
\end{tabular}\caption{LAHA projected interactions in the $d_{x^2-y^2}-$wave channel for  \mbox{k$_z =\pi$}.
The energies are in units of $U$.}
\end{center}
\begin{center}
\tabcolsep=0.10cm
\begin{tabular}{ccccccc}
$d_{x^2-y^2}$-wave & $\tilde{U}_{h_3h_3}$ & $\tilde{U}_{h_3e}$ & $\tilde{\alpha}_{h_3e}$ & $\tilde{U}_{ee}$ & $\tilde{\alpha}_{ee}$ & $\tilde{\beta}_{ee}$ \\
\hline
$J=0.0U$&	 0.01	&	 0.01	&	 0.71	&	 0.02	&	 0.71	&	 0.51\\
$J=0.125U$&	 0.01	&	 0.01	&	 0.71	&	 0.02	&	 0.71	&	 0.51\\
$J=0.25U$&	 0.01	&	 0.01	&	 0.71	&	 0.01	&	 0.71	&	 0.51\\
\hline
\end{tabular}\caption{LAHA projected interactions in the $d_{x^2-y^2}-$wave channel for \mbox{k$_z =0$}. The energies are in units of $U$.}
\end{center}
\end{table}
\end{center}

\begin{figure}[t!]
\includegraphics[width=1.0\textwidth]{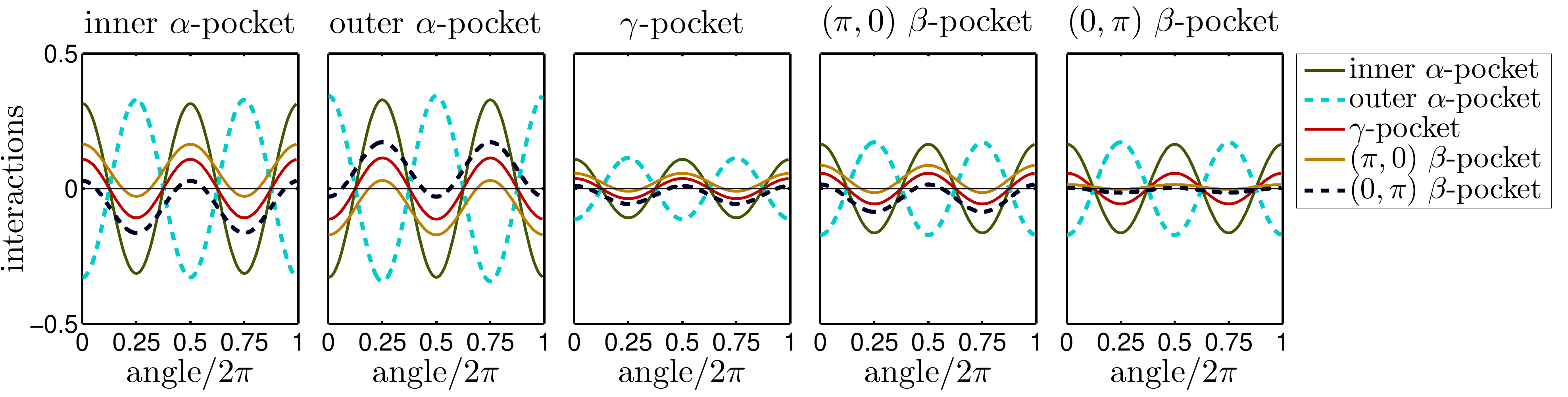}
\caption{ Behavior of the bare interactions $\Gamma^{d_{x^2-y^2}}{ij}({\bf k}_{F},{\bf k^\prime}_{F})$ on the Fermi surfaces of LiFeAs as obtained by LAHA procedure for $J=0.125U$. As in the $s$-wave case, we set ${\bf k}_F$ on the FS labeled by $i$ to be along $x$ direction  and vary ${\bf k}_F^\prime$ along each of the FSs labeled by $j$. The angle is measured relative to $k_x$.}
\label{fig:interactions_dx2y2}
\end{figure}

\end{widetext}
For $d_{x^2-y^2}$ pairing symmetry, the gaps on different FSs behave as
\begin{align*}
\begin{split}
\Delta_{h_1}(\phi)&=\Delta_{h_1}\cos{2\phi}	\\
\Delta_{h_2}(\phi)&=\Delta_{h_2}\cos{2\phi}	\\
\Delta_{h_3}(\phi)&=\Delta_{h_3}\cos{2\phi}	\\
\Delta_{e_1}(\theta)&=\Delta_{e}+\bar{\Delta}_{e}\cos{2\theta}\\
\Delta_{e_2}(\theta)&=-\Delta_{e}+\bar{\Delta}_{e}\cos{2\theta}\\
\end{split}
\end{align*}
Solving the set of 5 coupled equations for $k_z = \pi$ and the set of three coupled equations for $\Delta_{h_3}$,
$\Delta_{e}$, and $\bar{\Delta}_{e}$, we obtain the results shown in Table. VIII and Fig.6(central panel).
\begin{table}[hbtp]
\begin{minipage}[h]{0.33\textwidth}
\begin{center}
\tabcolsep=0.10cm
\begin{tabular}{cccccc}
					&\multicolumn{2}{c}{$k_z=\pi$}&	&\multicolumn{2}{c}{$k_z=0$}\\
\hline					
$\Delta_{h_1}$		&$+0.60$	&$+0.57$	&		&$$&$$	\\
$\Delta_{h_2}$		&$+0.45$	&$-0.00$	&		&$$&$$	\\	
$\Delta_{h_3}$		&$-0.62$	&$+0.00$	&		&$+0.10$	&$+0.36$\\
$\Delta_{e}$		&$-0.05$	&$-0.09$	&		&$-0.59$	&$+0.53$\\	
$\bar{\Delta}_{e}$	&$+0.22$	&$-0.81$	&		&$+0.80$	&$+0.76$\\
\hline
$\lambda$			&$0.00$		&$-0.00$	&		&$-0.00$	&$-0.07$\\
\hline
\end{tabular}\caption{$d_{x^2-y^2}$ solution for $J=0.125U$. As before, $\pm 0.00$ means that eigenvalue is positive (negative), but its magnitude is  smaller than $5 \times 10^{-3}$.}
\end{center}
\end{minipage}
\end{table}
Interestingly, the $d_{x^2-y^2}-$wave also changes sign between the two inner $\alpha-$hole pockets and also between the outer $\gamma-$hole pocket
and $\beta$-electron pockets.

\begin{widetext}

 \begin{figure}[t!]
 \includegraphics[width=1.0\textwidth]{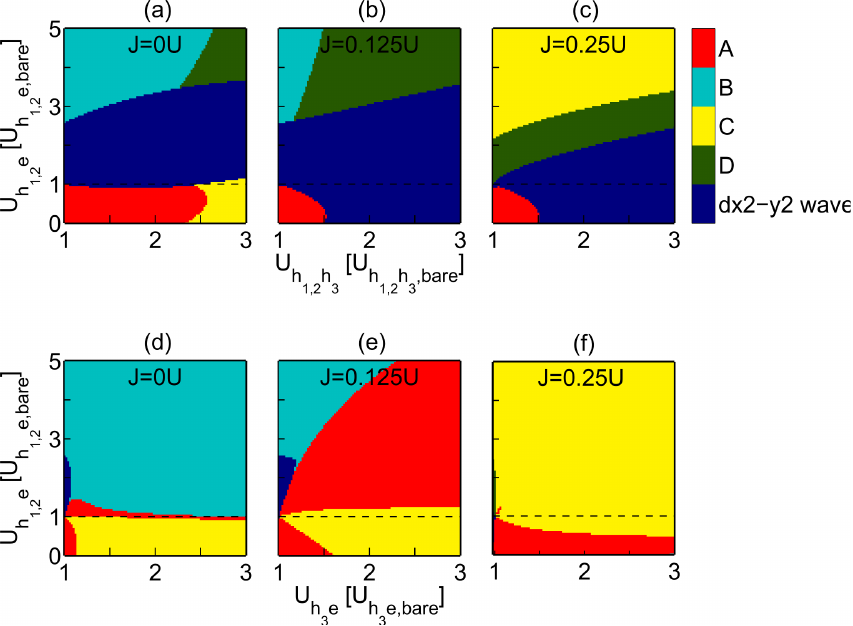}
 \caption{The phase diagram obtained by identifying the superconducting $s-$wave and $d$-wave gap structure for given parameters
 with the eienfunction corresponding to the largest eigenvalue of the linearized gap equation. We vary the coupling in a similar fashion for the $s-$wave and $d_{x^2-y^2}$-wave (the upper panel) either between two subsets I and II or between electron and hole pockets (the lower panel) . The bare (un-renormalized) values of parameters are the ones
 shown in Tables II for the $s-$wave and Tables VI, IX for the $d-$wave.}
 \label{phase_diagramwith_d-wave}
 \end{figure}

\end{widetext}

An important result is that for the bare parameters the leading eigenvalue for $d_{x^2-y^2}$-wave solution is very close to zero, like in the $s$-wave case.
As a result, $d_{x^2-y^2}$ and $s-$wave pairings are strong competitors. The competition holds when we  modify the corresponding interactions in $d-$wave and $s-$wave channels by the same amount. We show the results in Fig.\ref{phase_diagramwith_d-wave}. We see that in some regions $d_{x^2-y^2}$- state is the leading instability. Note, however, that if increase the the interactions between the two $\alpha$-pockets, the phase space for $d_{x^2-y^2}-$wave phase gets significantly reduced.
\begin{widetext}
\begin{center}
\begin{table}[hbtp]
\begin{center}
\tabcolsep=0.10cm
\begin{tabular}{ccccccccccc}
$d_{xy}$-wave & $\hat{U}_{h_1h_1}$ & $\hat{U}_{h_2h_2}$ & $\hat{U}_{h_3h_3}$ & $\hat{U}_{h_1h_2}$ & $\hat{U}_{h_1h_3}$ & $\hat{U}_{h_2h_3}$ & $\hat{U}_{h_1e}$ & $\hat{U}_{h_2e}$ & $\hat{U}_{h_3e}$ &  $\hat{U}_{ee}$ \\
\hline
$J=0.0U$& 	 0.54	&	 0.59	&	 0.04	&	 -0.57	&	 0.09	&	 -0.09	&	 -0.04	&	 0.04	&	 -0.02	&	 0.01\\
$J=0.125U$&  0.48	&	 0.51	&	 0.04	&	 -0.49	&	 0.08	&	 -0.08	&	 -0.04	&	 0.04	&	 -0.01	&	 0.01\\
$J=0.25U$& 	 0.41	&	 0.44	&	 0.03	&	 -0.42	&	 0.07	&	 -0.06	&	 -0.03	&	 0.03	&	 -0.01	&	 0.00\\
\hline
\end{tabular}\caption{LAHA projected interactions in the $d_{xy}-$wave channel for \mbox{k$_z =\pi$}. The energies are in units of $U$.}
\end{center}
\begin{center}
\tabcolsep=0.10cm
\begin{tabular}{cccccc}
$d_{xy}$-wave & $\hat{U}_{h_3h_3}$ & $\hat{U}_{h_3e}$ & $\hat{U}_{ee}$ \\
\hline
$J=0.0U$&	 0.01	&	 -0.01	&	 0.00\\
$J=0.125U$&	 0.01	&	 -0.01	&	 0.00\\
$J=0.25U$&	 0.00	&	 -0.00	&	 0.00\\
\hline
\end{tabular}\caption{LAHA projected interactions in the $d_{xy}-$wave channel for \mbox{k$_z =0$}. The energies are in units of $U$.}
\end{center}
\end{table}
\end{center}

\begin{figure}[t!]
 \includegraphics[width=1.0\textwidth]{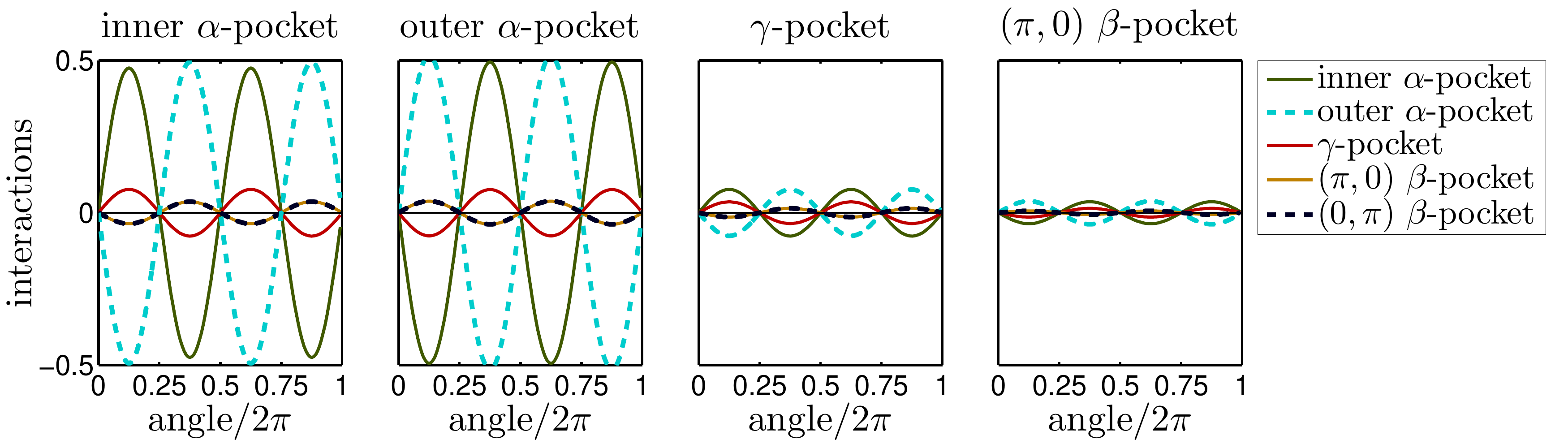}
     \caption{Behavior of the bare interactions $\Gamma^{d_{xy}}_{ij}({\bf k}_{F},{\bf k^\prime}_{F})$ on the Fermi surfaces of LiFeAs as obtained by LAHA procedure for $J=0.125U$. As in the $s$-wave and $d_{x^2-y^2}$-wave cases, we set ${\bf k}_F$ on the FS labeled by $i$ to be along $x$ direction and vary ${\bf k}_F^\prime$ along each of the FSs labeled by $j$. The angle is measured relative to $k_x$.}
     \label{fig:interactions_dxy}
 \end{figure}

\end{widetext}

For the interaction in the $d_{xy}$ channel we have only one set of eigenfunctions $\sin({4n +2} \theta_i)$ and $\sin({4n +2} \phi_i)$. Taking the leading $\sin 2\theta_i$ components we obtain
\begin{align*}
\Gamma^{d_{xy}}_{h_ih_j}(\phi,\phi^\prime)&=\hat{U}_{h_ih_j}\sin{\phi}\sin{\phi^\prime}\\
\Gamma^{d_{xy}}_{h_ie}(\phi,\theta)&=\hat{U}_{h_ie}\sin{\phi}\sin{\theta}\\
\Gamma^{d_{xy}}_{e_ie_j}(\theta,\theta^\prime)&=\hat{U}_{ee}\sin{\theta}\sin{\theta^\prime} 
\end{align*}
The coefficients ${\hat U}_{ij}$ are presented in Tables IX-X and in Fig. \ref{fig:interactions_dxy}

The gaps on different FSs are given by
\begin{align*}
\begin{split}
\Delta_{h_1}(\phi)&={\Delta}_{h_1}\sin{2\phi}	\\
\Delta_{h_2}(\phi)&={\Delta}_{h_2}\sin{2\phi}	\\
\Delta_{h_3}(\phi)&={\Delta}_{h_3}\sin{2\phi}	\\
\Delta_{e_1}(\theta)&={\Delta}_{e}\sin{2\theta}\\
\Delta_{e_2}(\theta)&={\Delta}_{e}\sin{2\theta}\\
\end{split}
\end{align*}
We present the results for the bare interactions and $J=0.125U$ Tables XI and Fig.6 (right panel)
\begin{table}[hbtp]
\begin{minipage}[h]{0.33\textwidth}
\begin{center}
\tabcolsep=0.10cm
\begin{tabular}{cccccc}
					&\multicolumn{2}{c}{$k_z=\pi$}&	&\multicolumn{2}{c}{$k_z=0$}\\
\hline					
$\Delta_{h_1}$		&$+0.56$	&$+0.58$	&		&$$&$$\\
$\Delta_{h_2}$		&$+0.55$	&$+0.53$	&		&$$&$$\\	
$\Delta_{h_3}$		&$-0.40$	&$+0.31$	&		&$+0.83$	&$+0.80$\\	
$\Delta_{e}$		&$-0.47$	&$+0.53$	&		&$+0.56$	&$-0.60$\\
\hline
$\lambda$			&$0.00$		&$-0.00$	&		&$-0.00$	&$-0.01$\\
\hline
\end{tabular}\caption{$d_{xy}$ solution, $J=.125U$}
\end{center}
\end{minipage}
\end{table}

We again see that for bare interactions, the leading eigenvalue is near zero. However, we found that, once we modify the system parameters in the same way as in Fig. \ref{phase_diagramwith_d-wave}, $d_{xy}$ state is subdominant to either $s$-wave or to $d_{x^2-y^2}$-state.

\section{Conclusion}
\label{sec:7_conclusion}

In this paper we used the tight-binding model, derived from {\it ab-initio} LDA calculations with hopping parameters extracted from the fit to ARPES experiments, and analyzed the structure of the pairing interaction and superconducting gap in LiFeAs.
We decomposed the pairing interaction for various $k_z$ cuts into $s-$ and $d$-wave components and analyzed the leading superconducting instabilities.
We focused on $s-$wave pairing as ARPES experiments ruled out d-wave superconductivity in  LiFeAs.

We find that, for bare interactions, the largest eigenfunction in the s-wave channel is zero to a very high accuracy, for all values of $J/U$ which we considered. In this situation, small changes in the intra-pocket and inter-pocket interactions due to renormalizations by high-energy fermions give rise to a variety of different gap structures, depending on which interactions get stronger by renormalizations.   We find four different configurations of the $s-$wave gap immediately below $T_c$:  the configuration in which the superconducting gap changes sign between two inner hole pockets and between the outer hole pocket and two electron pockets (state A); the one in which the gap changes sign between two electron pockets and three hole pockets (state B); the one in which the gap on the outer hole pocket differs in sign from the gaps on the other four pockets (state C); and the one in which the gaps on two inner hole pockets have one sign, and the gaps on the outer hole pockets and on electron pockets have different sign (state D). We associate the near-degeneracy between different s-wave states with two features. First, the pairing interaction almost decouples  between two subsets, one of which consists of the outer hole pocket and two electron pockets, which are quasi-2D and are made mostly out of $d_{xy}$ orbital, and the other consists of the two inner hole pockets, which are quasi-3D and are made mostly out of $d_{xz}$ and $d_{yz}$ orbitals.  Second, bare inter-pocket and intra-pocket interactions within each subset are nearly equal.  Different s-wave states emerge depending on whether the renormalized interactions increase the attraction  within each subset or the coupling between the subsets.
We discuss the phase diagram and the experimental probes to determine the structure of the superconducting gap in LiFeAs. We argue that the state A  with opposite sign of the gaps on the two inner hole pockets has the best overlap with ARPES data.

We found that the four configurations gradually transform into each other at $T_c$, upon changing the system parameters (except for one special case). However,   below $T_c$,  the transformation from one state to the other is not necessary a continuous one and may occur via an intermediate state, or a set of states, in which the phases of the gaps on different pockets differ by less than $\pi$. In these $s+is$ states time-reversal symmetry is spontaneously broken. Whether any of these states is realized in LiFeAs at a low $T$ remains to be seen, but the search for potential time-reversal symmetry broken superconductivity in LiFeAs is clearly called for.

We acknowledge helpful discussions and collaborations with L. Benfatto, A. Kreisel, P.
Hirschfeld, J-P Hu, G. Kotliar, S. Maiti, Y. Matsuda, I. Mazin, R. Moessner,  P.Thalmeier, and Y. Wang. The work was supported by
the DFG under priority programme SPP 1458 and German Academic Exchange Service (DAAD PPP USA No. 57051534). J.K.~acknowledges support from the Studienstiftung des deutschen Volkes, the IMPRS Dynamical Processes in Atoms, Molecules and Solids, DFG within GRK1621. S.V.B., V.B.Z. and B.B. acknowledge support under Grants No. ZA 654/1-1, No. BO1912/2-2, and No. BE1749/13. A.V.C. is supported by the Office of BES,  US DOE
 grant \#DE-FG02-ER46900.

.

\end{document}